\documentclass[conference]{IEEEtran}
\usepackage{cite}
\usepackage{amsmath,amssymb,amsfonts}
\usepackage{algorithmic}
\usepackage{graphicx}
\usepackage{textcomp}
\usepackage{xcolor}
\usepackage{fancyhdr}
\usepackage[hyphens]{url}

\def\BibTeX{{\rm B\kern-.05em{\sc i\kern-.025em b}\kern-.08em
    T\kern-.1667em\lower.7ex\hbox{E}\kern-.125emX}}

\fancypagestyle{firstpage}{
  \fancyhf{}

  \fancyfoot[C]{\thepage}
}

\title{QADAM: Quantization-Aware DNN Accelerator Modeling for Pareto-Optimality}
\author{Ahmet Inci\textsuperscript{1}, Siri Garudanagiri Virupaksha\textsuperscript{1}, Aman Jain\textsuperscript{1}, Venkata Vivek Thallam\textsuperscript{1}, \\ Ruizhou Ding\textsuperscript{1}, Diana Marculescu\textsuperscript{1,2} \\ 
Carnegie Mellon University\textsuperscript{1}, The University of Texas at Austin\textsuperscript{2}\\
 {\tt\small \{ainci, sgarudan, amanj, vthallam, rding, dianam\}@andrew.cmu.edu}

} 

\begin{document}
\maketitle
\thispagestyle{firstpage}
\pagestyle{plain}


\begin{abstract}
As the machine learning and systems communities strive to achieve higher energy-efficiency through custom deep neural network (DNN) accelerators, varied bit precision or quantization levels, there is a need for design space exploration frameworks that incorporate quantization-aware processing elements (PE) into the accelerator design space while having accurate and fast power, performance, and area models. 
In this work, we present \textit{QADAM}, a highly parameterized quantization-aware power, performance, and area modeling framework for DNN accelerators. Our framework can facilitate future research on design space exploration and Pareto-efficiency of DNN accelerators for various design choices such as bit precision, PE type, scratchpad sizes of PEs, global buffer size, number of total PEs, and DNN configurations. 
Our results show that different bit precisions and PE types lead to significant differences in terms of performance per area and energy. Specifically, our framework identifies a wide range of design points where performance per area and energy varies more than $5 \times$ and $35 \times$, respectively. 
We also show that the proposed lightweight processing elements (LightPEs) consistently achieve Pareto-optimal results in terms of accuracy and hardware-efficiency. With the proposed framework, we show that LightPEs achieve on par accuracy results and up to $5.7 \times$ more performance per area and energy improvement when compared to the best INT16 based design. 
\end{abstract}

\section{Introduction}
Deep neural networks (DNNs) have achieved remarkable accomplishments across various applications ranging from image recognition \cite{EfficientNet}, object detection \cite{EfficientDet}, to natural language processing \cite{Devlin2019BERTPO}. However, the increasing model size and computational cost of these models become a challenging task for on-device machine learning (ML) endeavours due to the stringent performance per area and energy constraints of the edge devices. To this end, while machine learning practitioners focus on model compression techniques \cite{han2015deep_compression,ruizhou2018lightnn,Chin2020CVPR}, computer architects investigate hardware architectures to overcome the energy-efficiency problem and improve the overall system performance \cite{eyeriss,aly2015next,han2016eie,chen2017dataflow,Shao2019SimbaSD, inci2018asbd,DeepNVM,inci2021tcad,inci2020architectural,inci2021cross,inci2022qappa}.

As computing community hits the limits on consistent performance scaling for traditional architectures, there has been a rising interest on enabling on-device machine learning through custom DNN accelerators. As we deeply care about performance per area and energy-efficiency from a hardware point of view, tailored DNN accelerators have shown significant improvements when compared to CPUs and GPUs \cite{eyeriss,tpu,Parashar2017SCNNAA,tetris,tpuv4}. To better understand the trade-offs of various architectural design choices and DNN workloads, there is a need for a design space exploration framework that can rapidly iterate over various designs and generate power, performance, and area (PPA) results. To this end, in this work we present \textit{QADAM}, a quantization-aware power, performance, and area modeling framework for DNN accelerators.

\begin{figure}[t]
  \centering
 \includegraphics[width=0.5\textwidth]{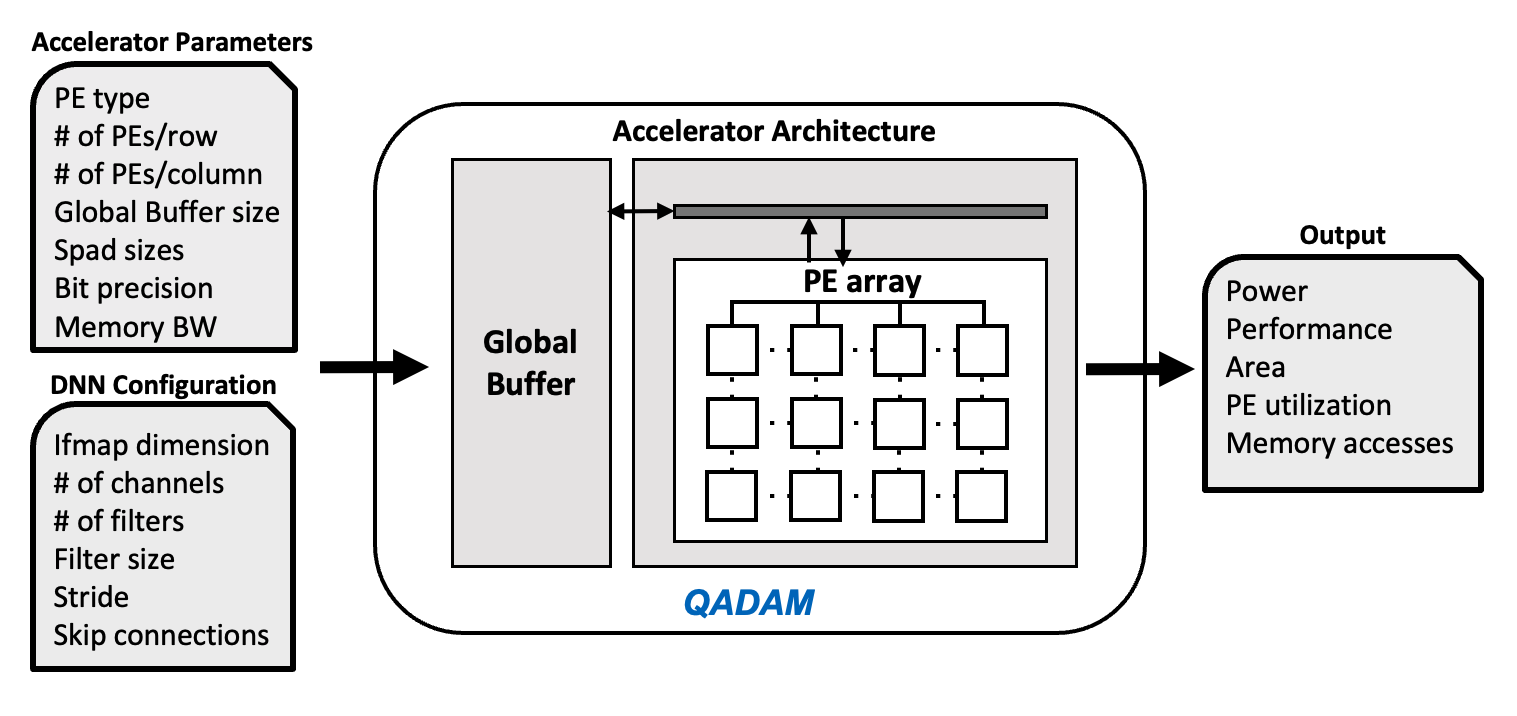}
 \caption{{Schematic depicting \textit{QADAM} framework, with accelerator parameters and DNN configuration as inputs. The framework takes in accelerator parameters and layer-wise DNN configurations and generates power, performance, area results, and statistics on hardware utilization and memory accesses.}}
 \vspace{-2mm}
 \label{fig:qappa}
\end{figure}

\begin{figure*}[t]
  \centering
 \includegraphics[width=0.8\textwidth]{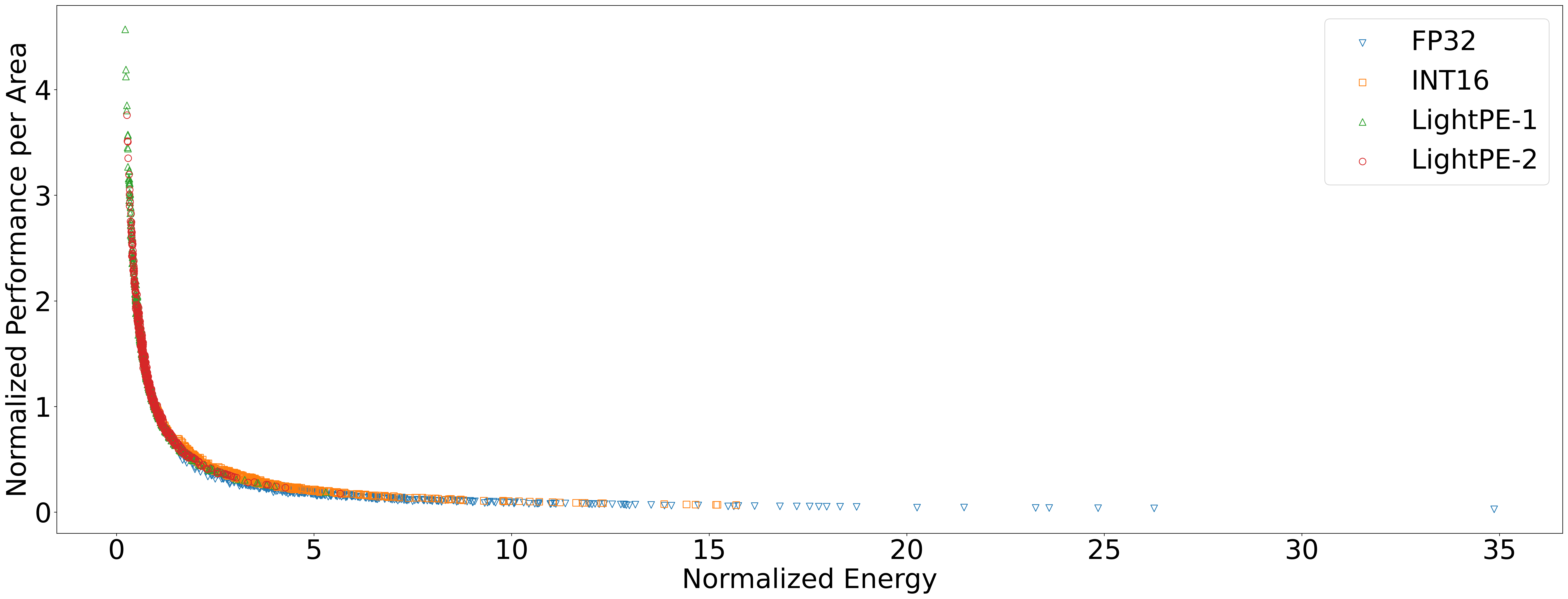}
 \caption{{Different PE types and bit precision lead to significant differences in performance per area and energy. Therefore, there is a need for a design space exploration framework that incorporates quantization-aware processing elements and rapidly iterate over various designs.}}
 \vspace{-2mm}
\label{fig:divergence}
\end{figure*}

This work makes the following contributions: 
\begin{itemize}
    \item We present QADAM, a quantization-aware power, performance, and area modeling framework for DNN accelerators. Our framework can enable future research on design space exploration of DNN accelerators for various design choices such as bit precision, processing element types, scratchpad sizes of processing elements, global buffer size, device bandwidth, number of total processing elements in the design, and DNN workloads. 
    \item  Our framework provides power, performance, and area results not just for a single hardware design point but for a range of different hardware designs as opposed to prior art \cite{qi17paleo,cai2017neuralpower}. Thus, it can be used to analyze trade-offs of various architectural design choices and DNN workloads at the same time to achieve Pareto-optimal design points in terms of accuracy and hardware-efficiency metrics such as performance per area and energy. 
    
\end{itemize}

The rest of the paper is organized as follows. In Section II, we present a literature review on power and runtime models for CNNs and design space exploration frameworks for hardware accelerators. 
In Section III, we describe the architectural details of the \textit{QADAM} framework and the details of our methodology for power, performance, and area modeling of DNN accelerators. In Section IV, we show experimental results demonstrating the efficiency of \textit{QADAM}'s PPA models and the efficacy of lightweight processing elements to conventional designs in terms of performance per area and energy through a suite of case studies. 
Finally, Section V concludes the paper by summarizing the results.

\section{Related Work}
Prior art has proposed runtime and energy models for DNN workloads \cite{cai2017neuralpower,qi17paleo,marculescu2018modelling}. However, these models have been implemented specifically for GPU platforms and thus they create an important limitation for a design space exploration of hardware architectures and potentially hardware and machine learning model co-design opportunities \cite{zhou2021rethinking,gupta2020acceleratoraware,codesign_vikas}. On the other hand, systems community has proposed tools and simulation methodologies for accelerator design. For example, SCALE-Sim \cite{samajdar2018scale} is a cycle accurate, systolic-array based DNN accelerator simulator. Similarly, Aladdin \cite{Shao2014AladdinAP} is a pre-RTL power and performance accelerator simulator. Although these tools help to perform preliminary analysis on the design space for accelerators in different aspects, they do not incorporate specialized quantization-aware processing elements and they do not generate RTL output of the chosen design based on the input hardware configuration which is an important impediment for enabling deployment of DNNs onto edge devices, as the actual deployment of the hardware design takes significant amount of engineering effort.

\section{Methodology}
In this section, we first explain the implementation details and architectural components of our \textit{QADAM} framework, as depicted in Figure~\ref{fig:qappa}. Next, we detail the lightweight processing elements (LightPE) that we implemented in our framework to provide a specialized processing element (PE) type for quantized DNN models. Finally, we explain our power, performance, and area modeling and design space exploration methodology.

\begin{figure*}[t]
  \centering
 \includegraphics[width=0.24\textwidth]{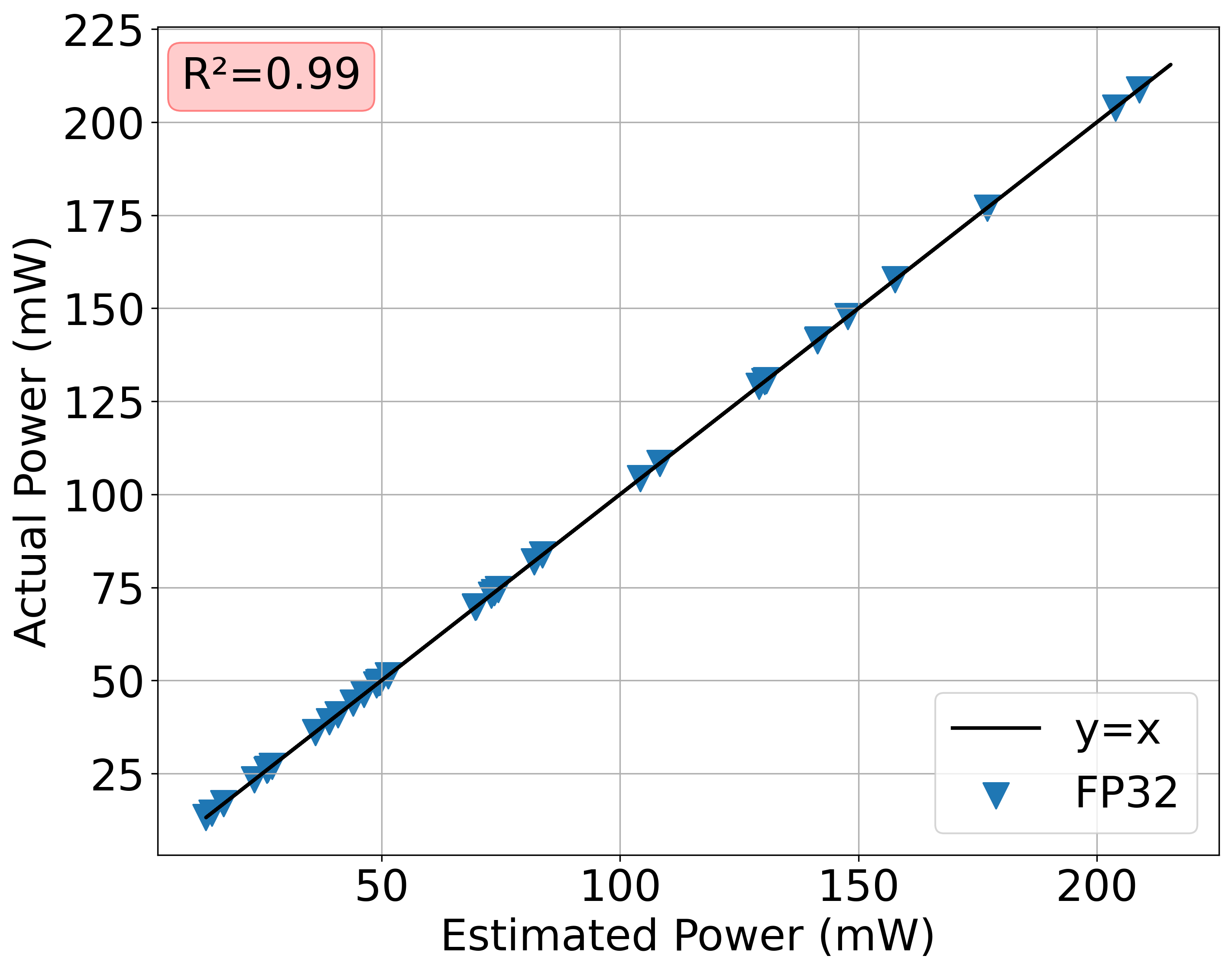}
  \includegraphics[width=0.24\textwidth]{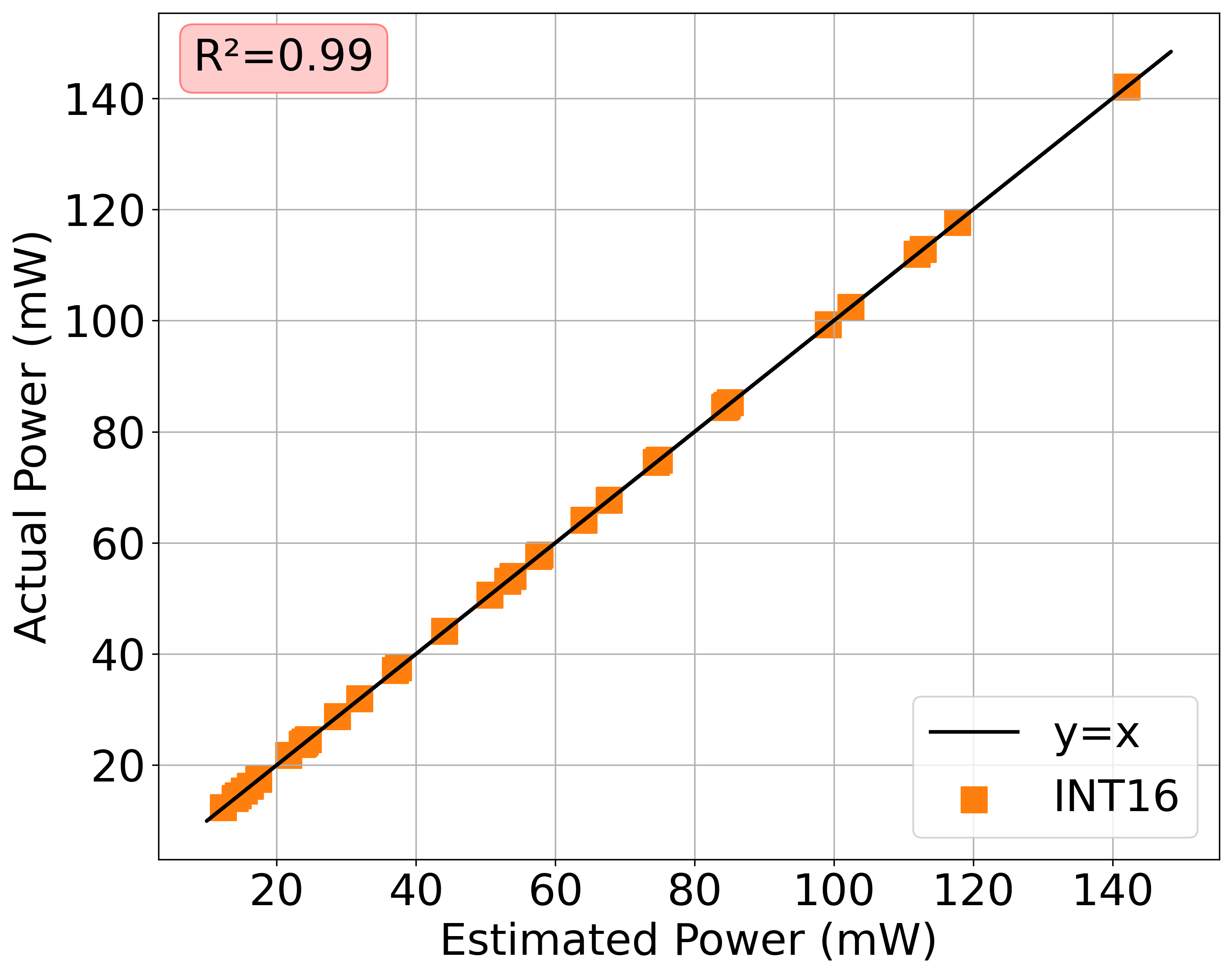}
  \includegraphics[width=0.24\textwidth]{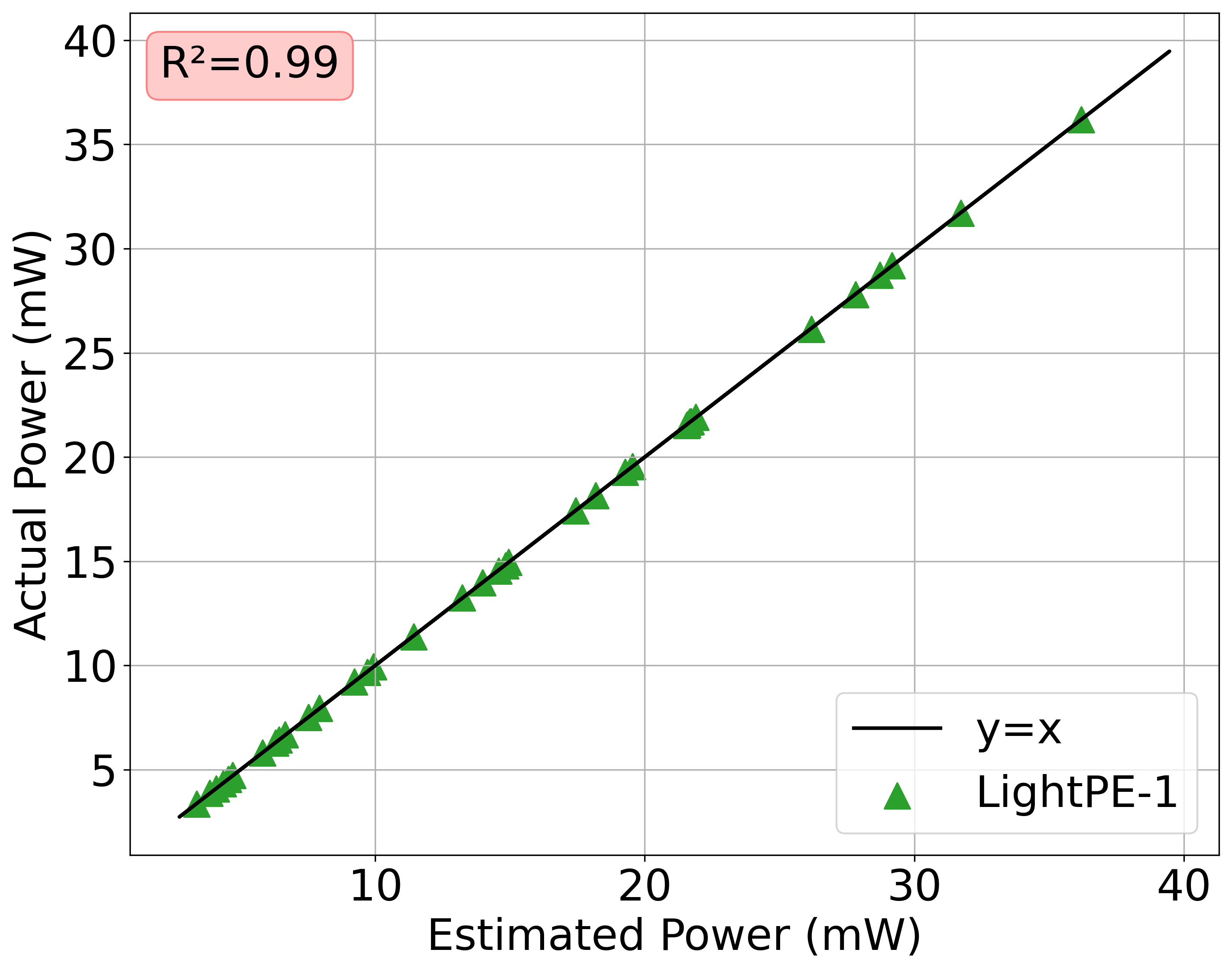}
    \includegraphics[width=0.24\textwidth]{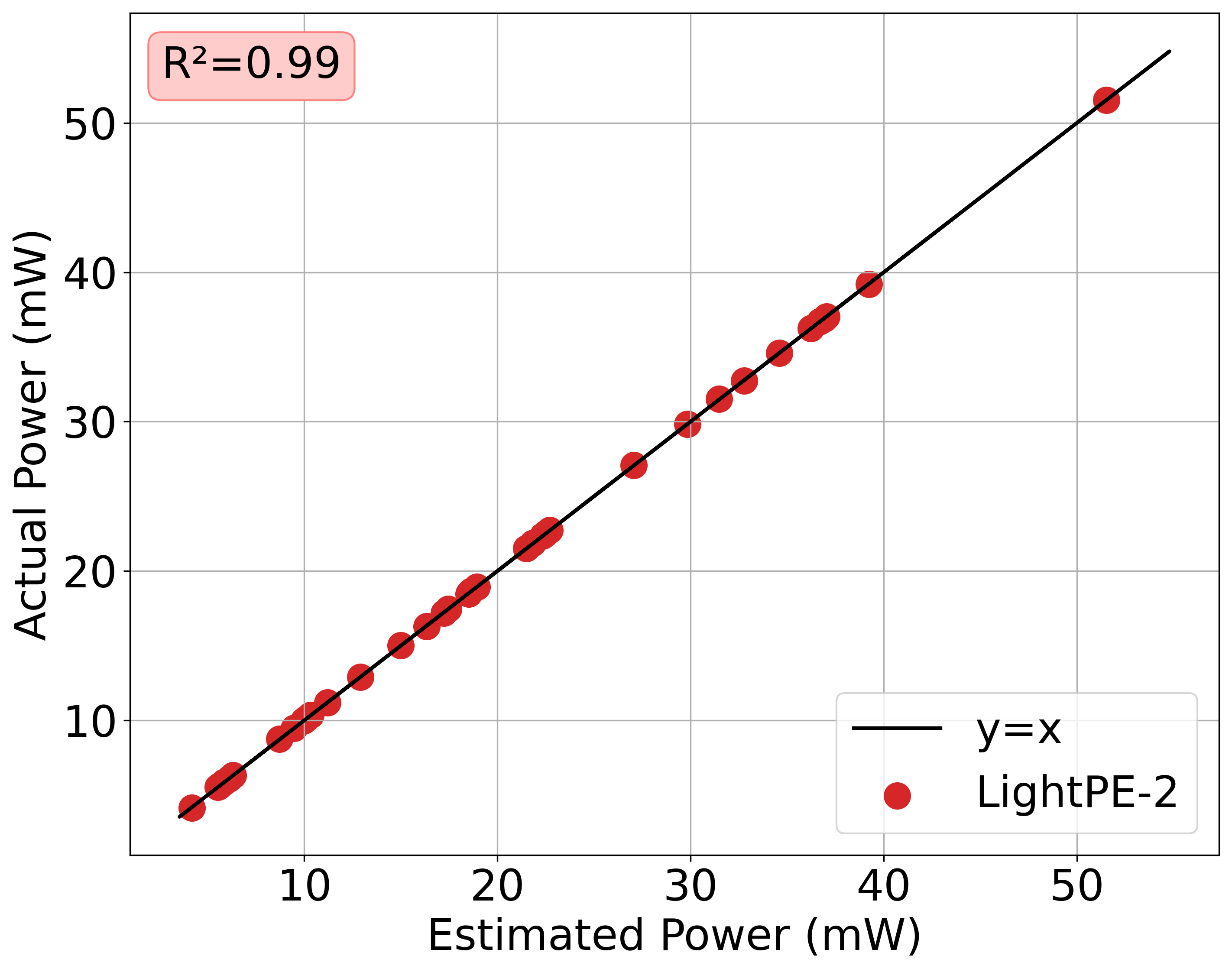}
    \\
     \includegraphics[width=0.24\textwidth]{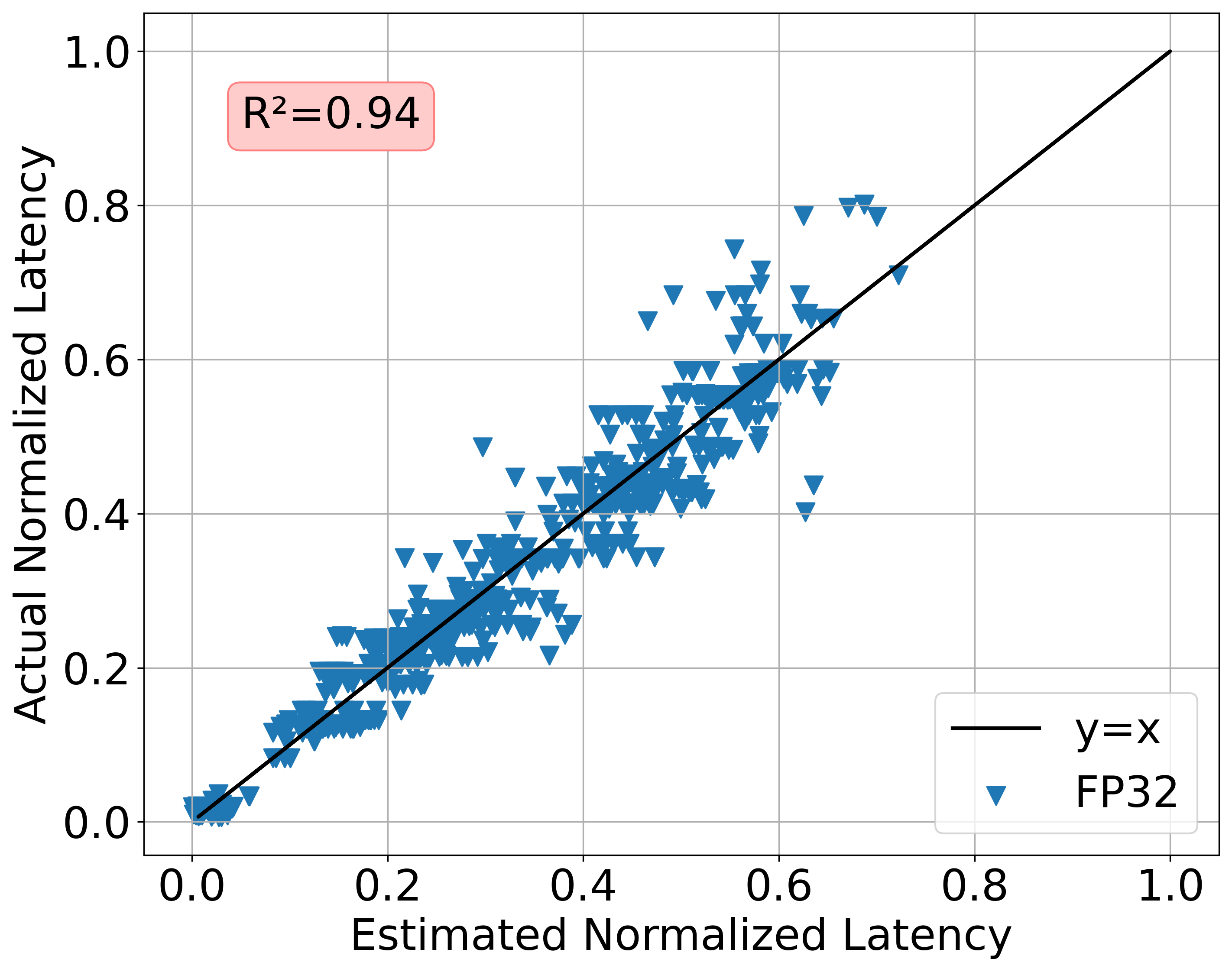} 
  \includegraphics[width=0.24\textwidth]{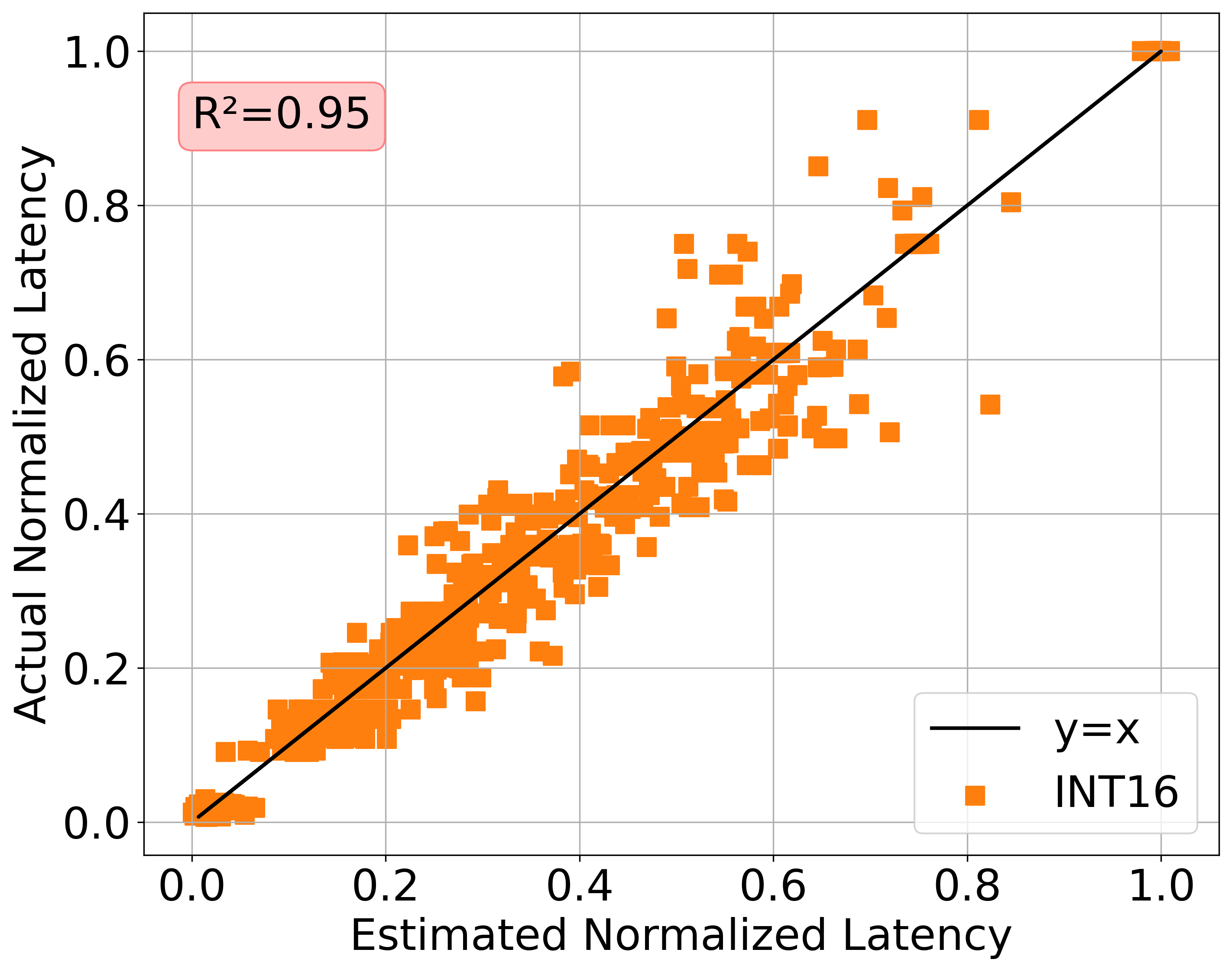}
  \includegraphics[width=0.24\textwidth]{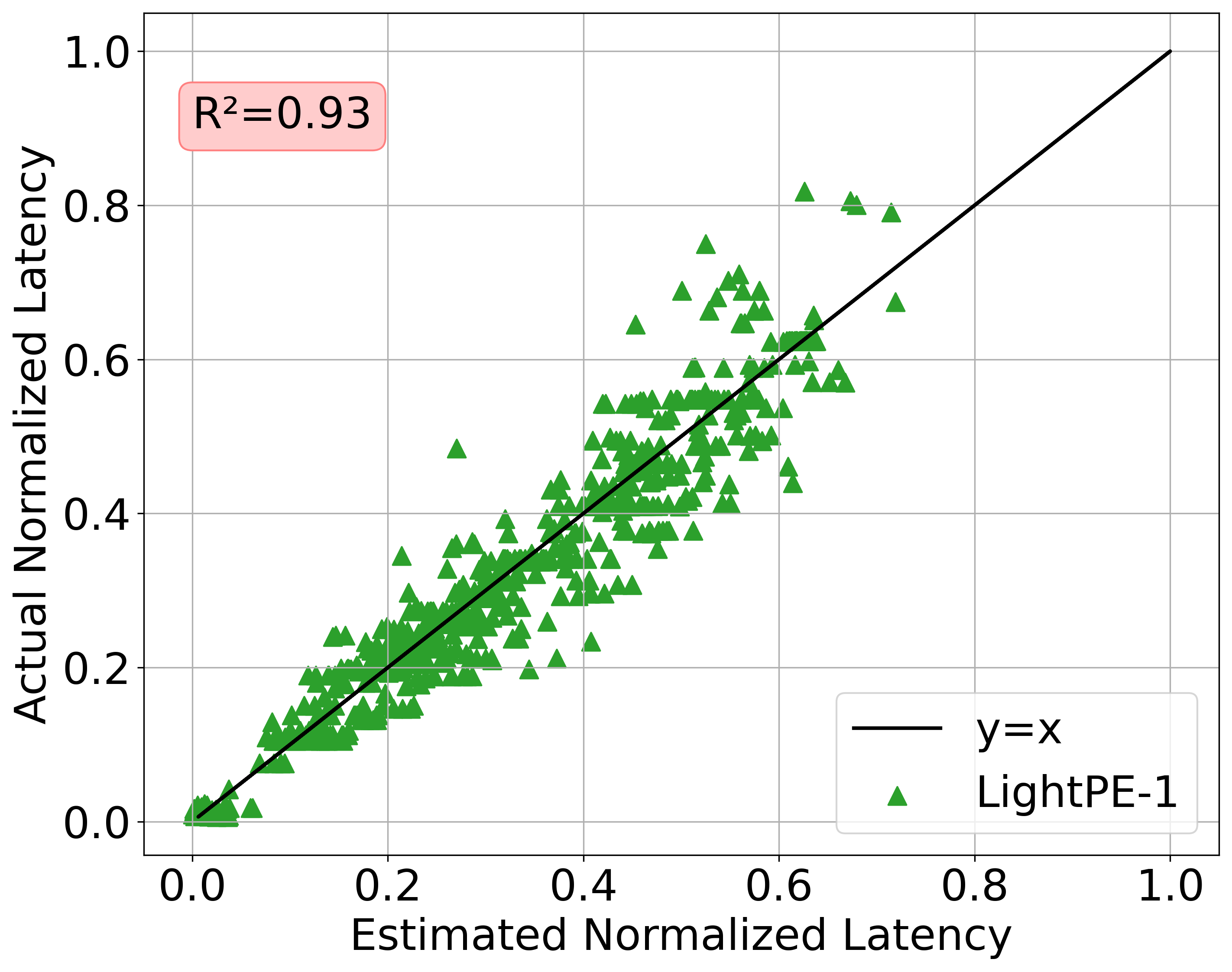}
    \includegraphics[width=0.24\textwidth]{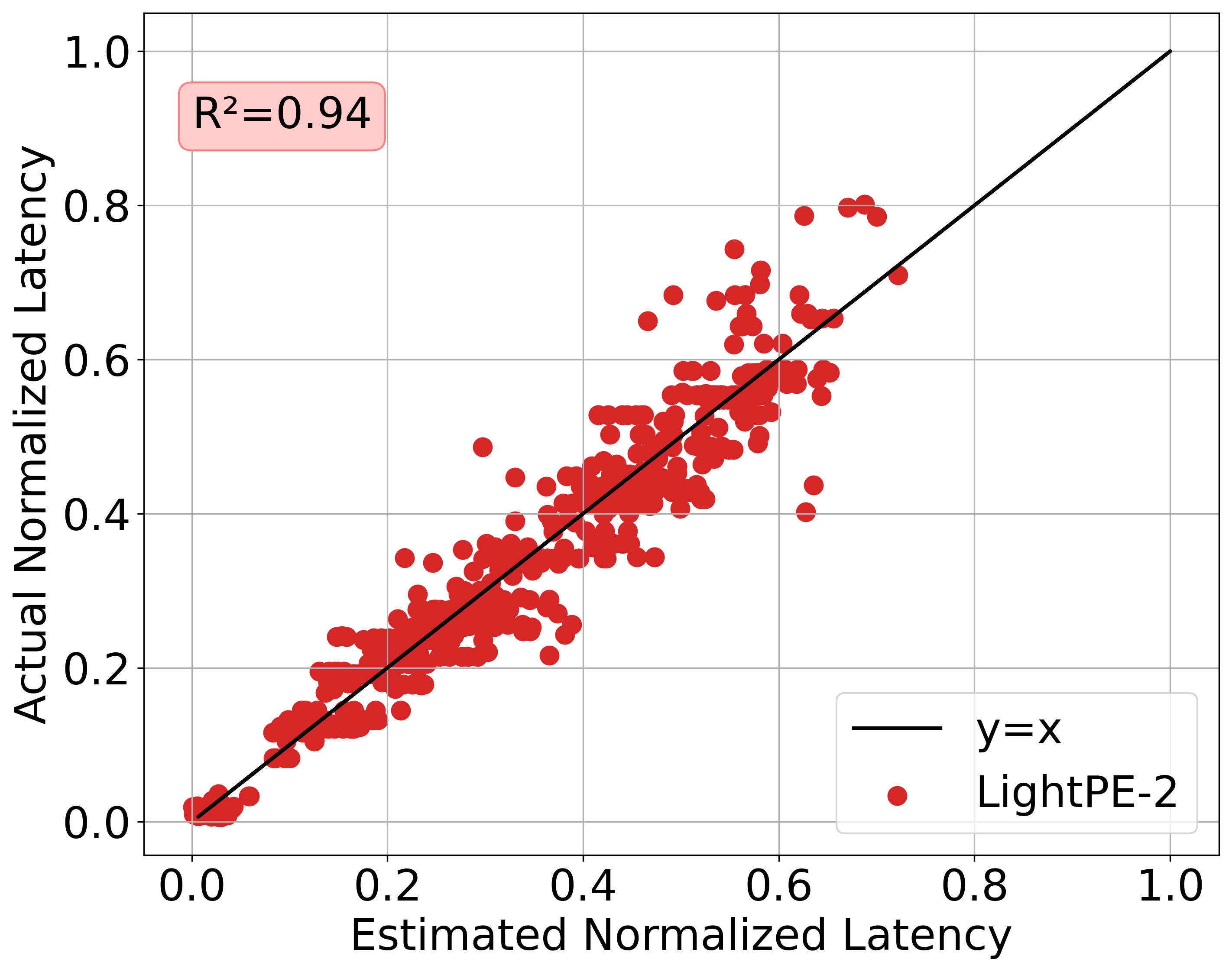}
    \\
 \includegraphics[width=0.24\textwidth]{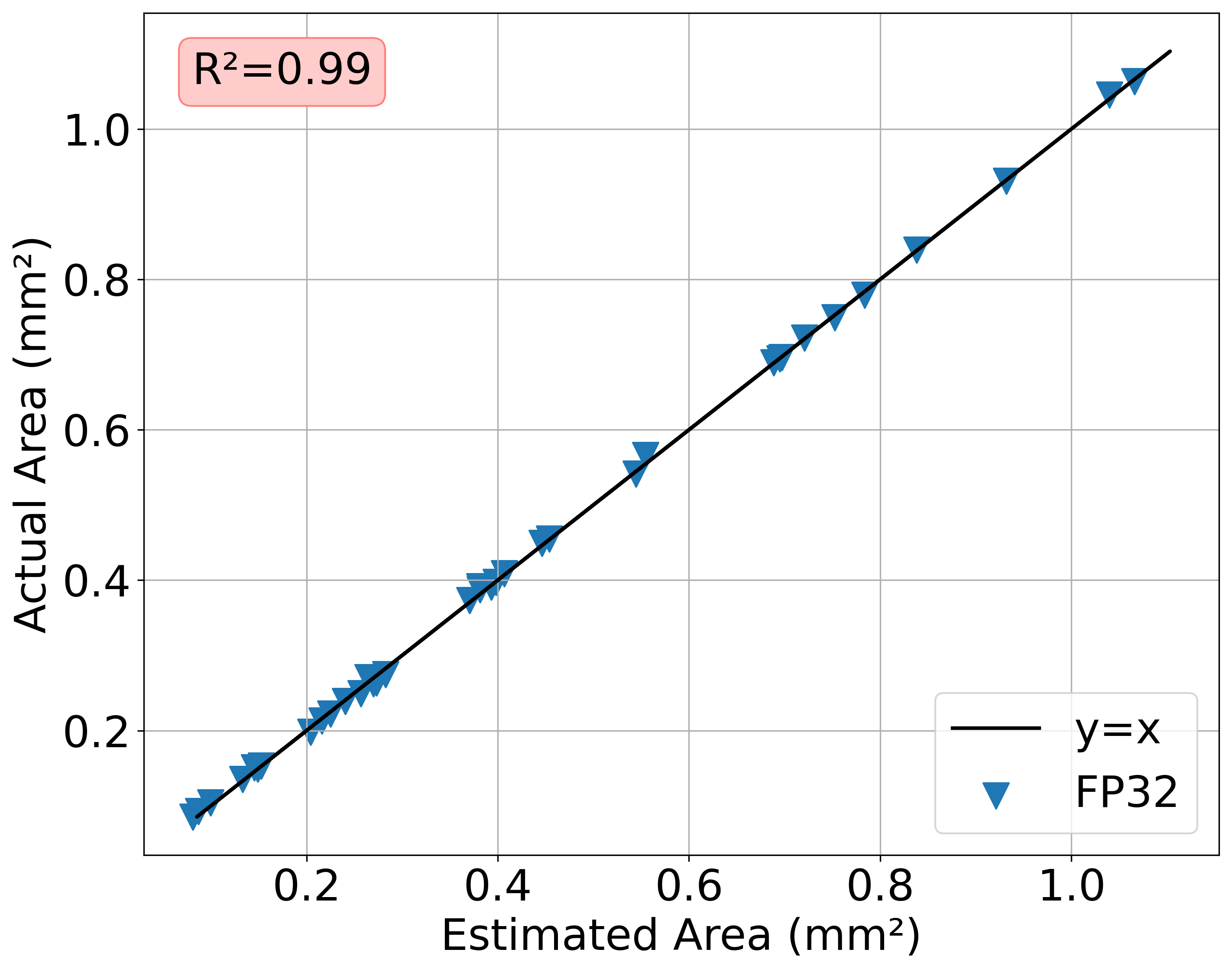}
  \includegraphics[width=0.24\textwidth]{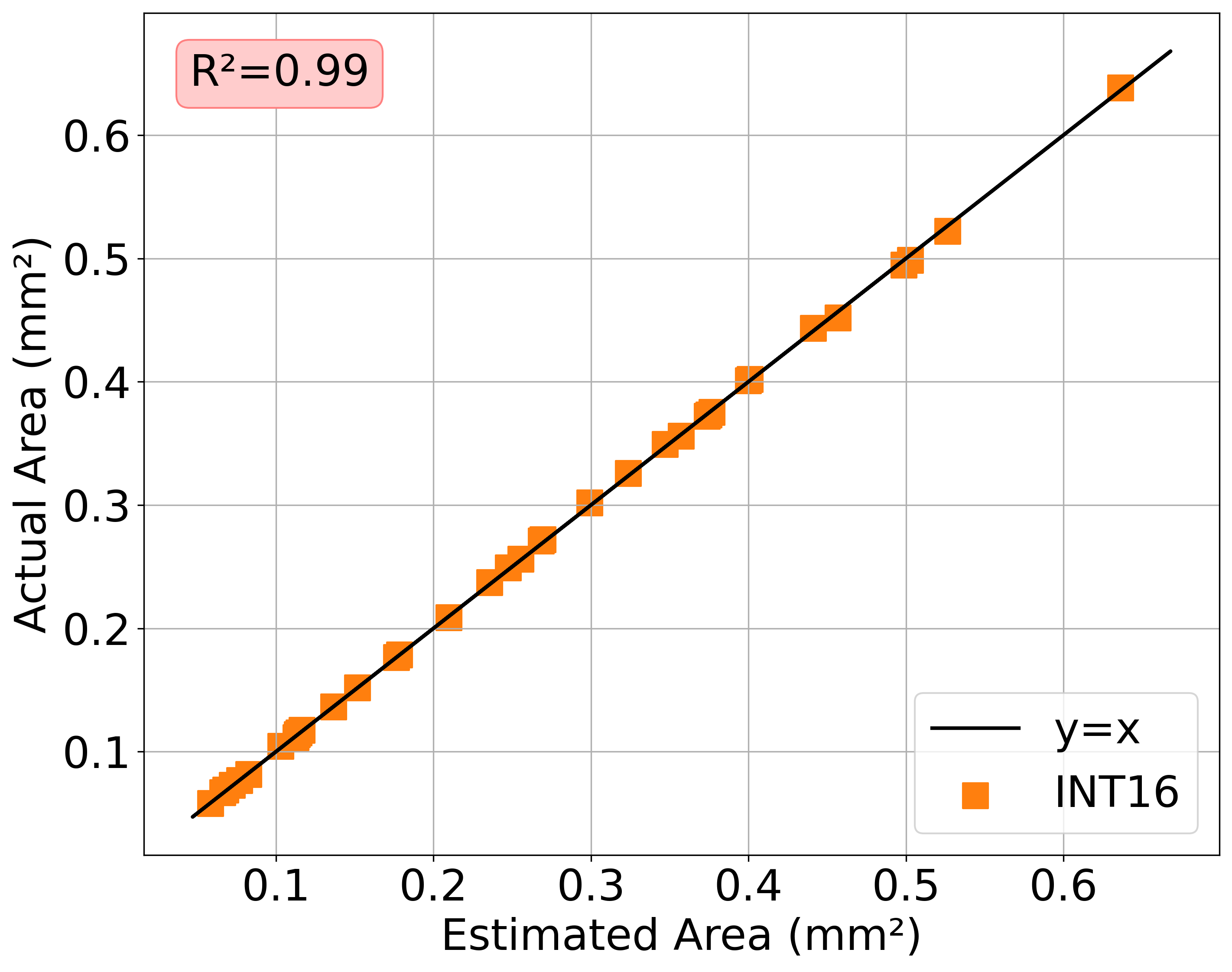}
  \includegraphics[width=0.24\textwidth]{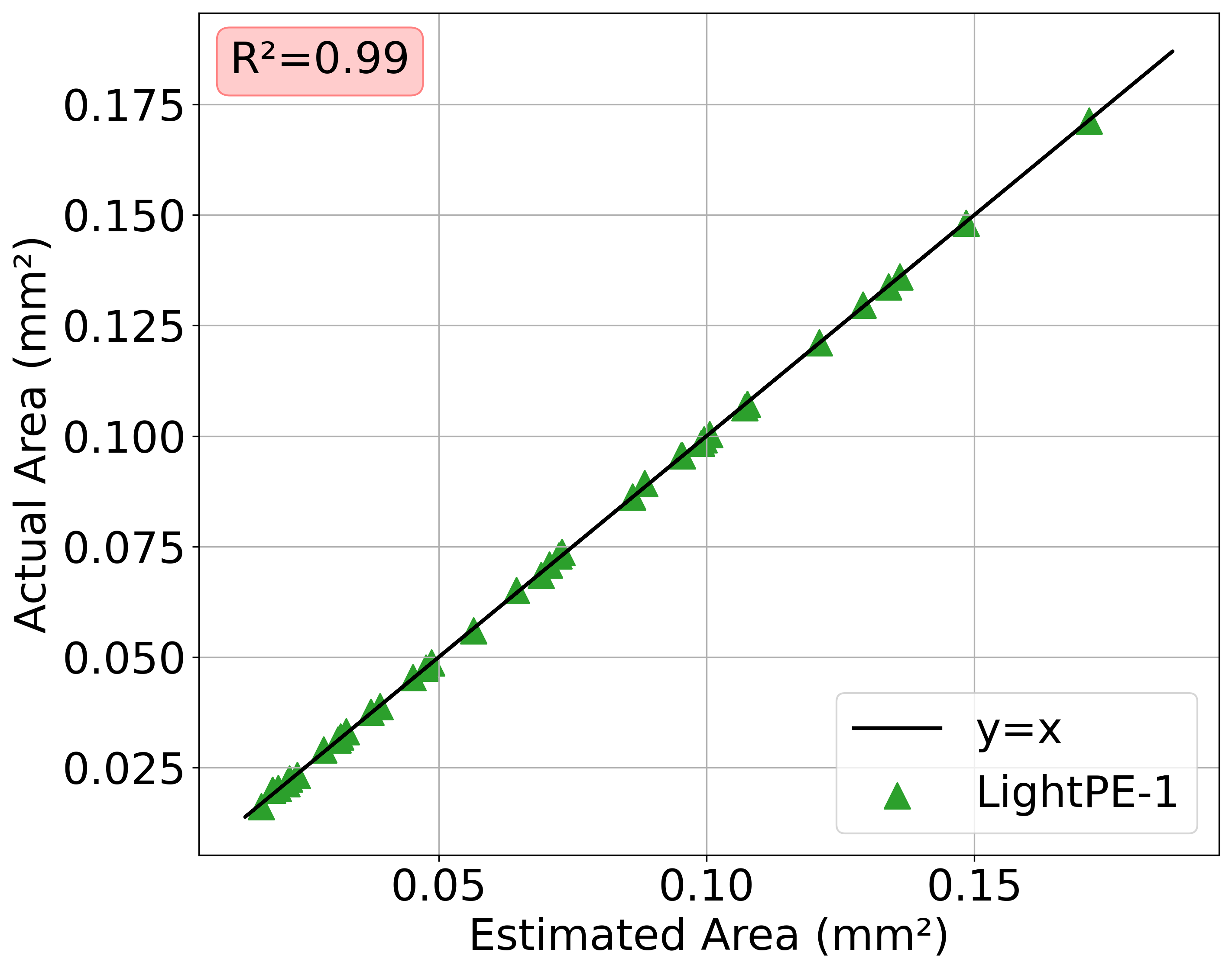}
    \includegraphics[width=0.24\textwidth]{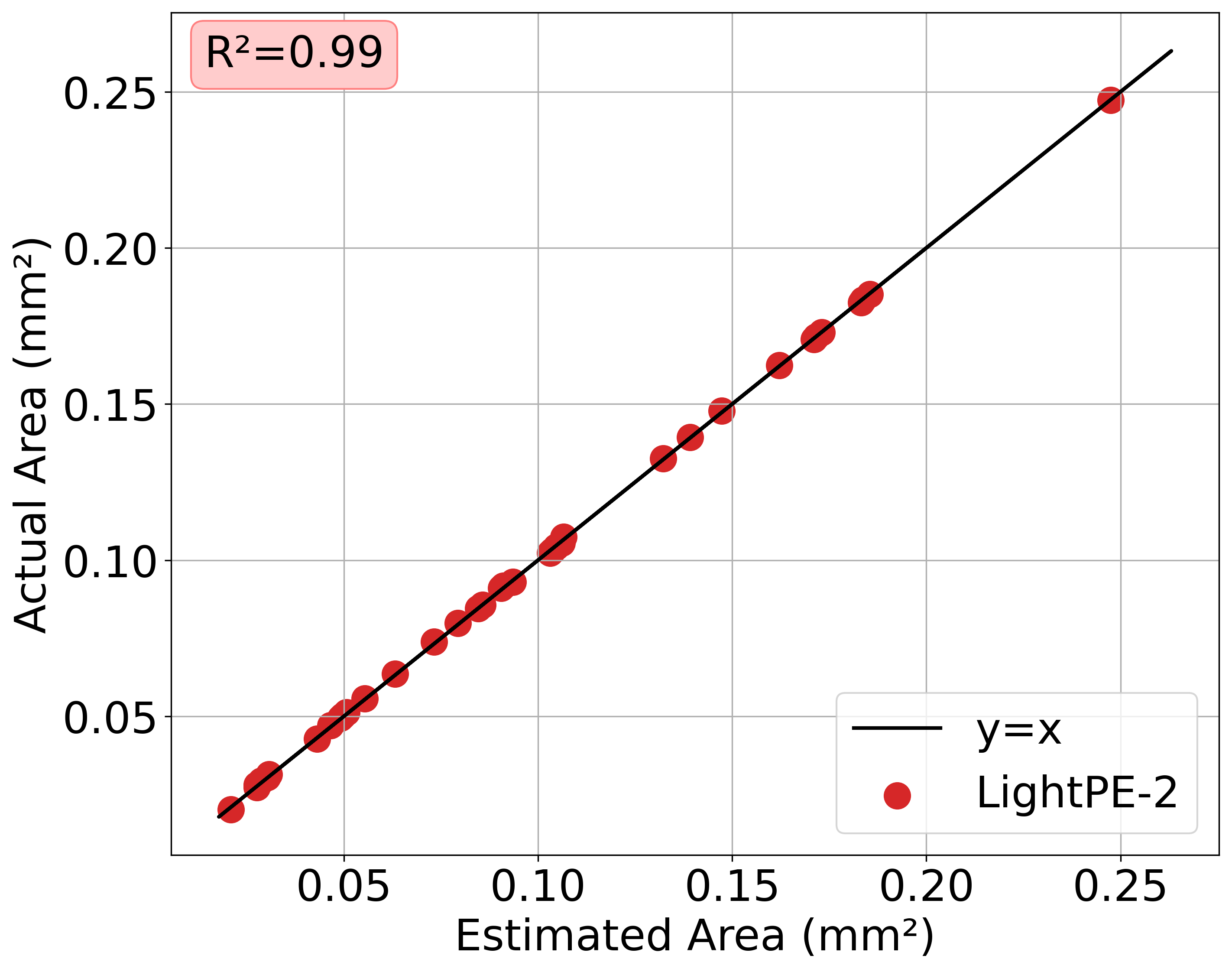}
    
 \caption{{Power (top chart), performance (middle chart), and area (bottom chart) estimation results for various processing element types such as FP32, INT16, LightPE-1, and LightPE-2. Each data point corresponds to a different hardware configuration that can be achieved by using the corresponding processing element type. As it can be seen, the proposed polynomial model agrees closely with the actual values extracted from the synthesis tools.}}
  \vspace{-2mm}
 \label{fig:model}
\end{figure*}

\subsection{QADAM Framework}

To enable comprehensive design space exploration for DNN accelerators for on-device machine learning, we implemented \textit{QADAM}, a highly parameterized spatial-array based DNN accelerator framework in RTL. Our framework enables hardware designers and machine learning practitioners to rapidly iterate over various accelerator designs and DNN configurations and better understand trade-offs of different architectural components of the design for dizzying requirements of deploying machine learning models to edge devices. Moreover, hardware designers can also use the automatically generated RTL code to follow the design synthesis flow. 

As depicted in Figure~\ref{fig:qappa}, \textit{QADAM} framework is based on spatial-array based accelerators and utilizes row stationary dataflow which has been demonstrated to optimize the data movement in the storage hierarchy \cite{eyeriss}. \textit{QADAM} features a set of processing elements organized as a 2D array and a global buffer that stores input feature maps, filters, and activations. The number of PEs in each dimension can be tuned for different power, performance, and area requirements. In each PE, there are input feature map, filter, and partial sum scratchpads and a multiply-accumulate (MAC) unit which can be chosen between a conventional MAC unit and a shift-add unit based on the desired bit precision. Each of these architectural components can be tuned in a flexible and automated manner to perform a comprehensive design space exploration for on-device edge accelerators. 

\subsection{Lightweight Processing Elements (LightPE)} 

To enrich the design space of hardware accelerators and achieve a better Pareto-frontier in terms of performance per area and energy-efficiency perspectives, we include LightPE implementations in our framework. LightPEs utilize 8 bits for activations and 4 bits and 8 bits for weights for LightPE-1 and LightPE-2 designs, respectively. As 4 bit and 8 bit quantization techniques for on-device machine learning became prevalent in various computing platforms, we provide these specialized quantization-aware PE types in our \textit{QADAM} framework to help hardware designers to enrich their design space and find better Pareto-frontiers.

Besides their low-precision benefits such as reducing the storage requirements, LightPEs also replace the multiplications with more energy and area-efficient one shift or a limited number of shifts and add operations \cite{ruizhou2018lightnn}. Therefore, they also achieve significant power and area gains when compared to full-precision 32 bit floating point (FP32) and 16 bit integer (INT16) based designs with only slight accuracy degradation \cite{ruizhou2018lightnn}. As a result, LightPEs provide an enriched design space for hardware designers and machine learning practitioners to analyze various trade-offs between accuracy and performance per area and energy. 
To this end, Figure~\ref{fig:divergence} shows that different PE types and precision lead to significant differences in terms of performance per area and energy. These results also reinforce the need for a design space exploration framework that incorporates quantization-aware hardware.

\subsection{Power, Performance, and Area Modeling} 

To build our quantization-aware power, performance, and area models, we use various hardware and DNN configurations. Specifically, to cover this comprehensive design space of hardware accelerators, we run experiments by varying global buffer size, number of PEs per row and column in the 2D PE array, bit precision, and PE type (FP32, INT16, LightPE-1, and LightPE-2). Within each PE, we also vary individual scratchpad sizes for input feature map, filter, and partial sum. 

We use Synopsys Design Compiler and the open-source FreePDK45 which is a commonly used process design kit \cite{freepdk} to synthesize our designs to obtain power, area, and initial timing results. We use Synopsys VCS RTL simulator to perform functional verification and collect timing information for various DNN configurations such as VGG-16 \cite{vgg16}, ResNet-20, ResNet-34, ResNet-50, and ResNet-56 \cite{resnet50} that are implemented in our testbenches. 
After collecting power, area, and timing results, we use polynomial regression models and model selection techniques based on \textit{k}-fold cross validation \cite{kfold} to tune the model parameters and fit the model. 

\vspace{-1mm}

\section{Results}
\vspace{-2mm}

\begin{figure*}[t]
  \centering
 \includegraphics[width=0.32\textwidth]{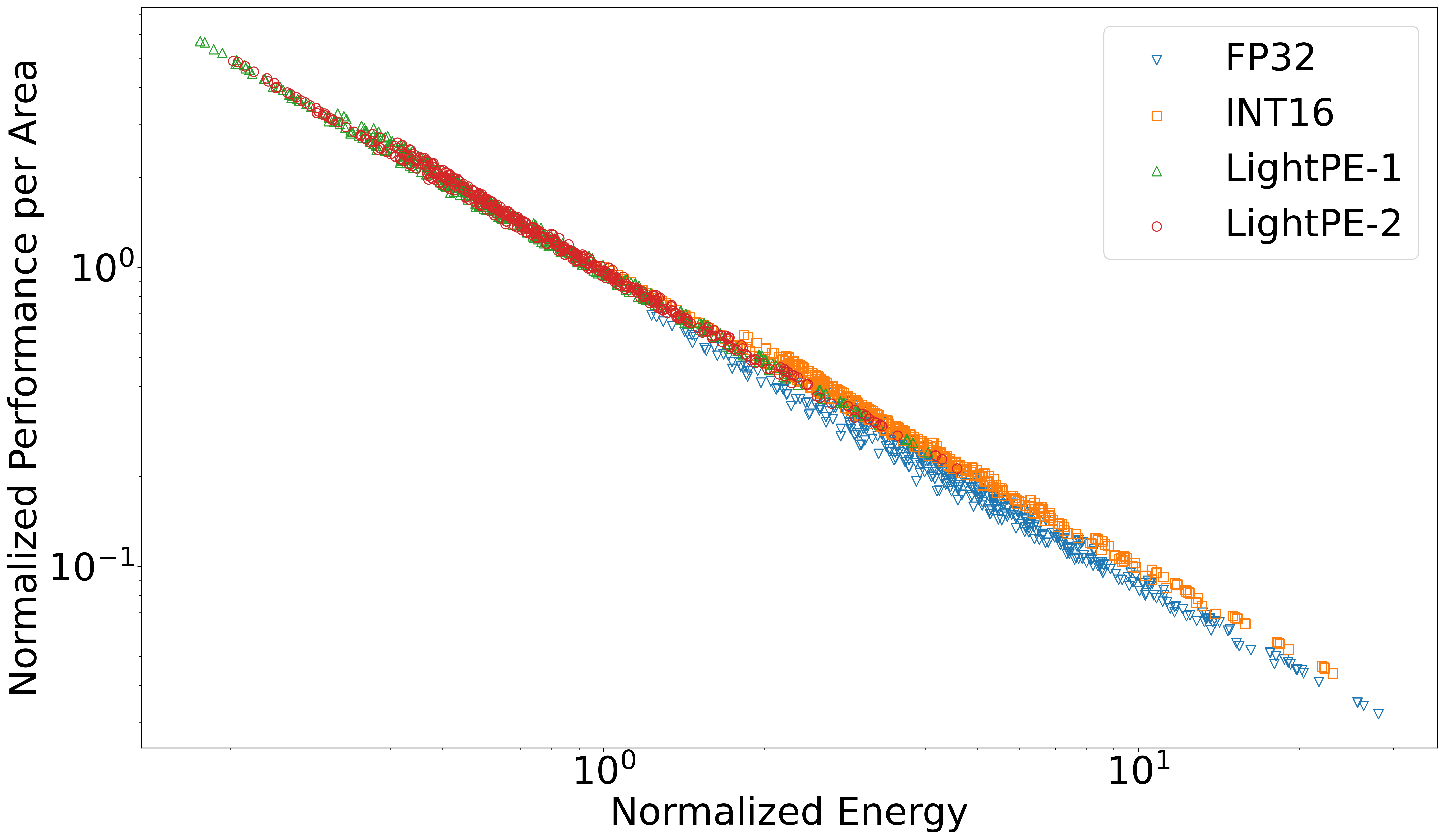}
  \includegraphics[width=0.32\textwidth]{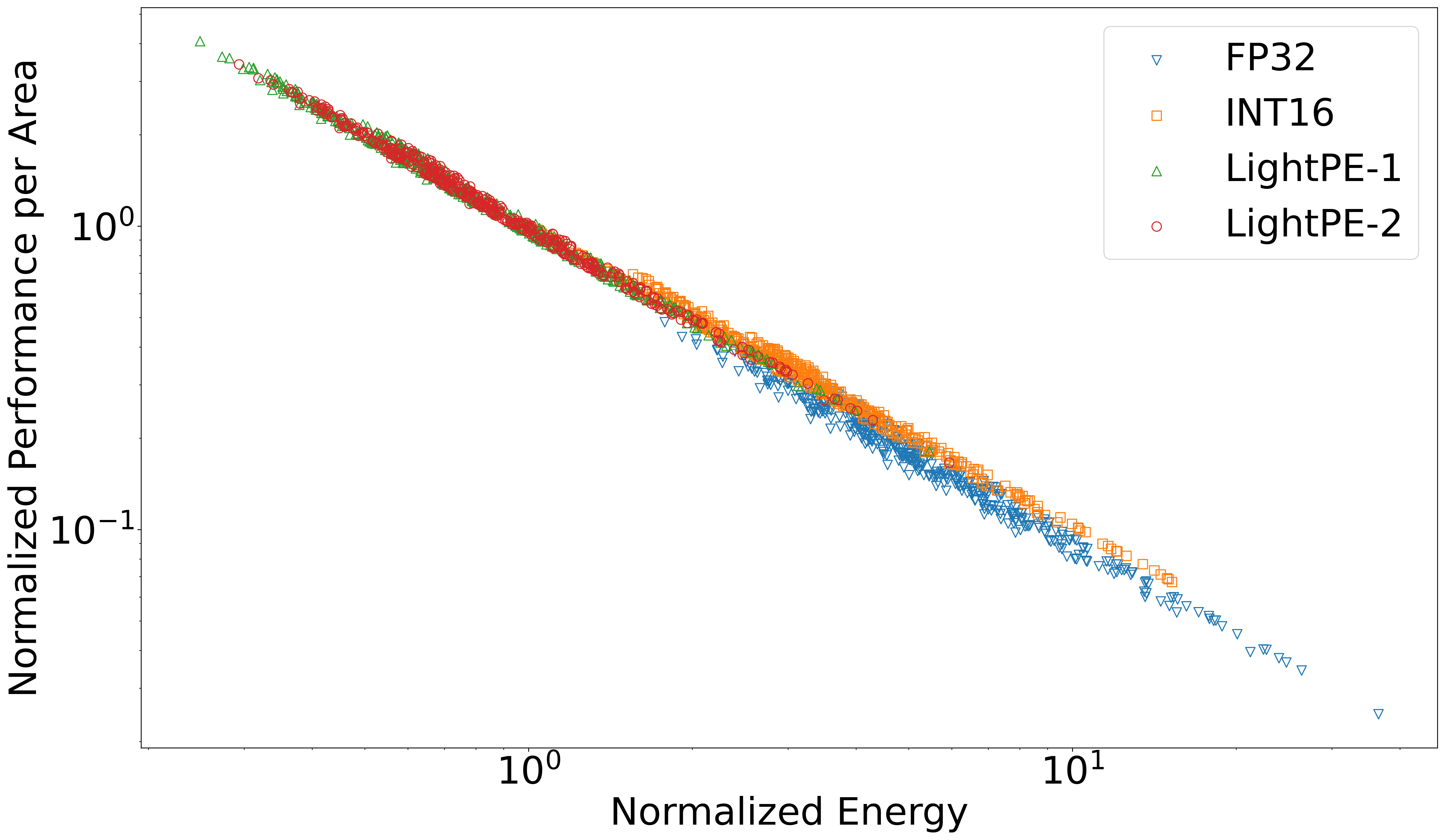}
  \includegraphics[width=0.32\textwidth]{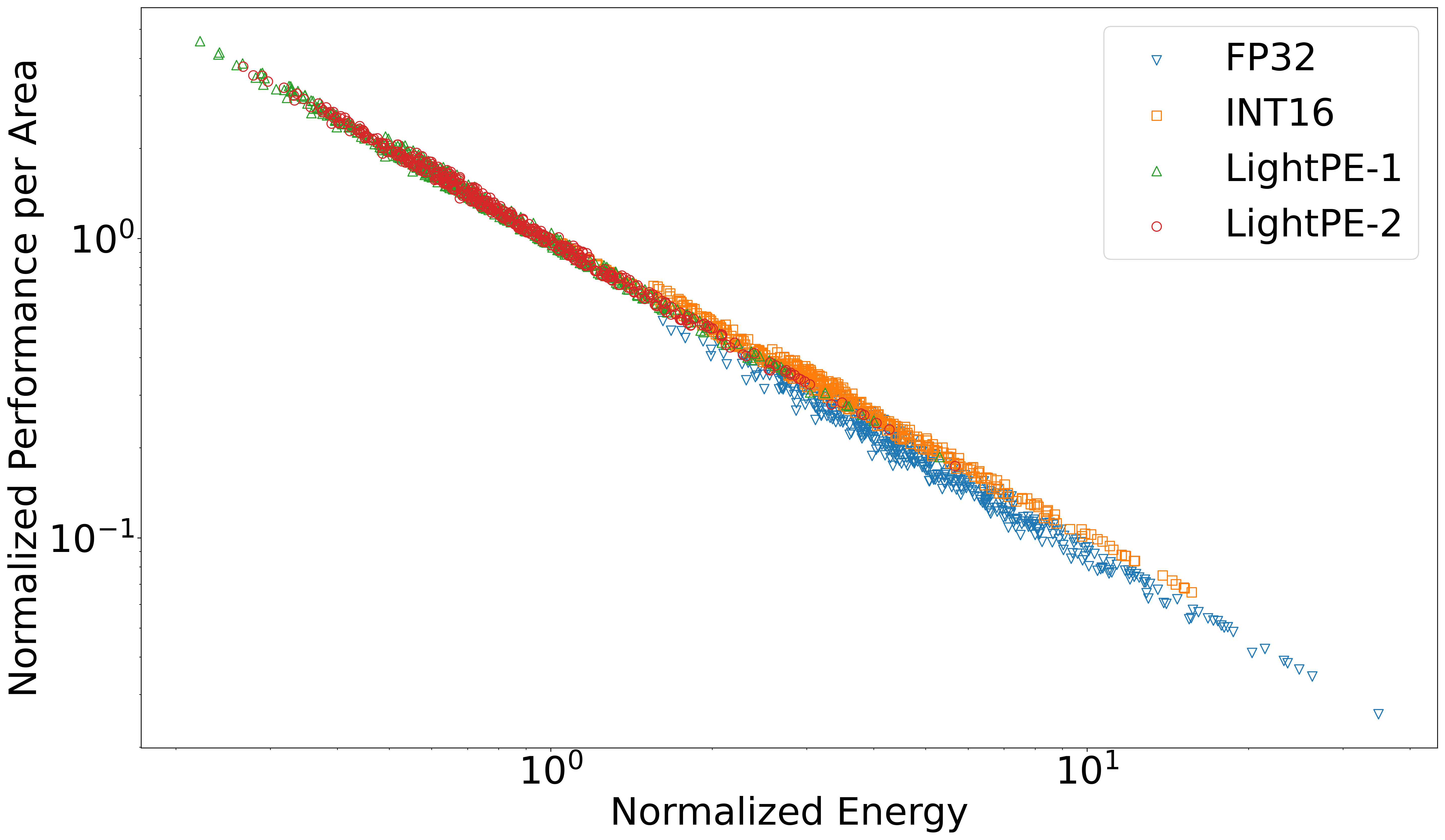}
    \\
     \includegraphics[width=0.32\textwidth]{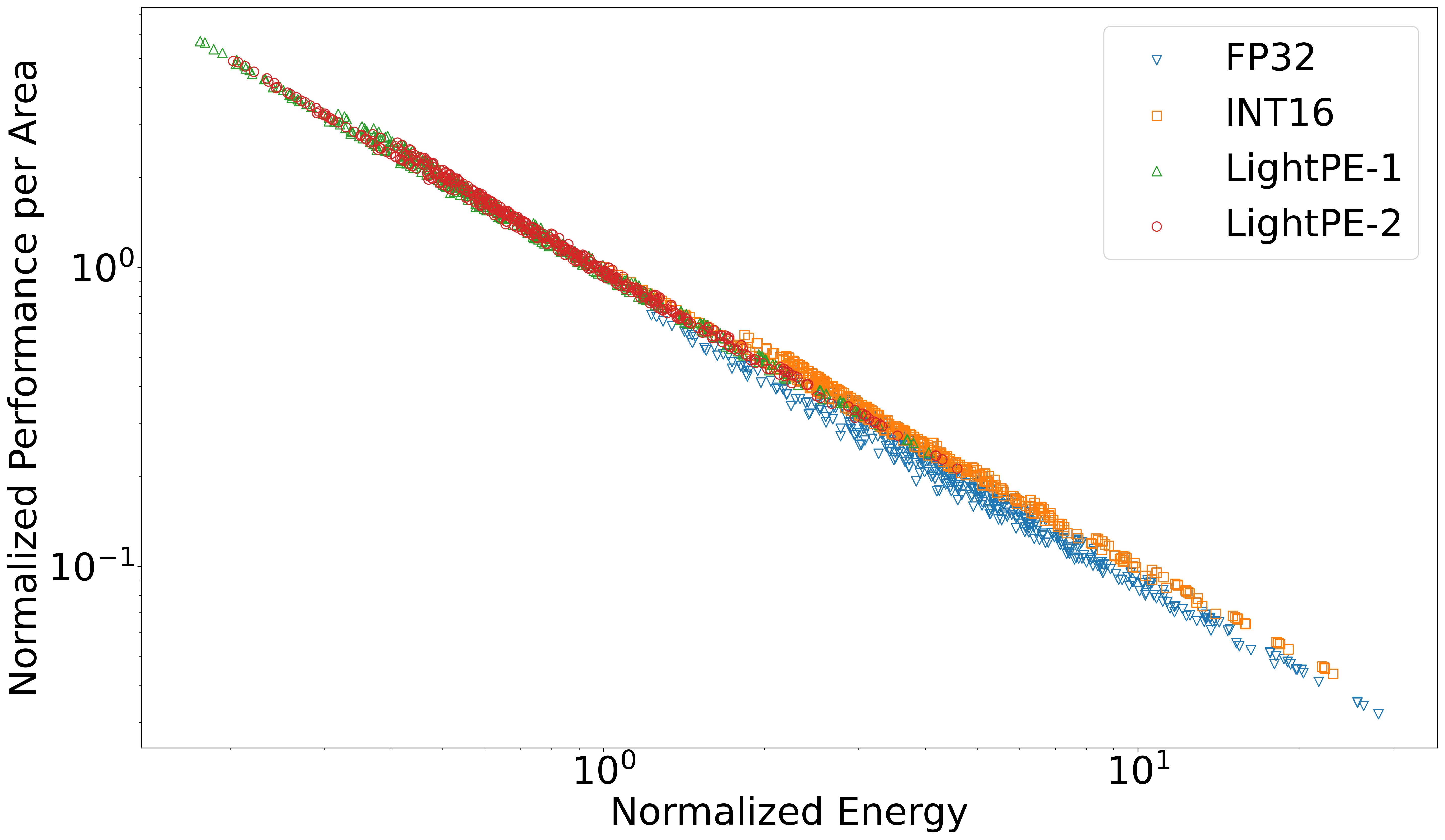}
  \includegraphics[width=0.32\textwidth]{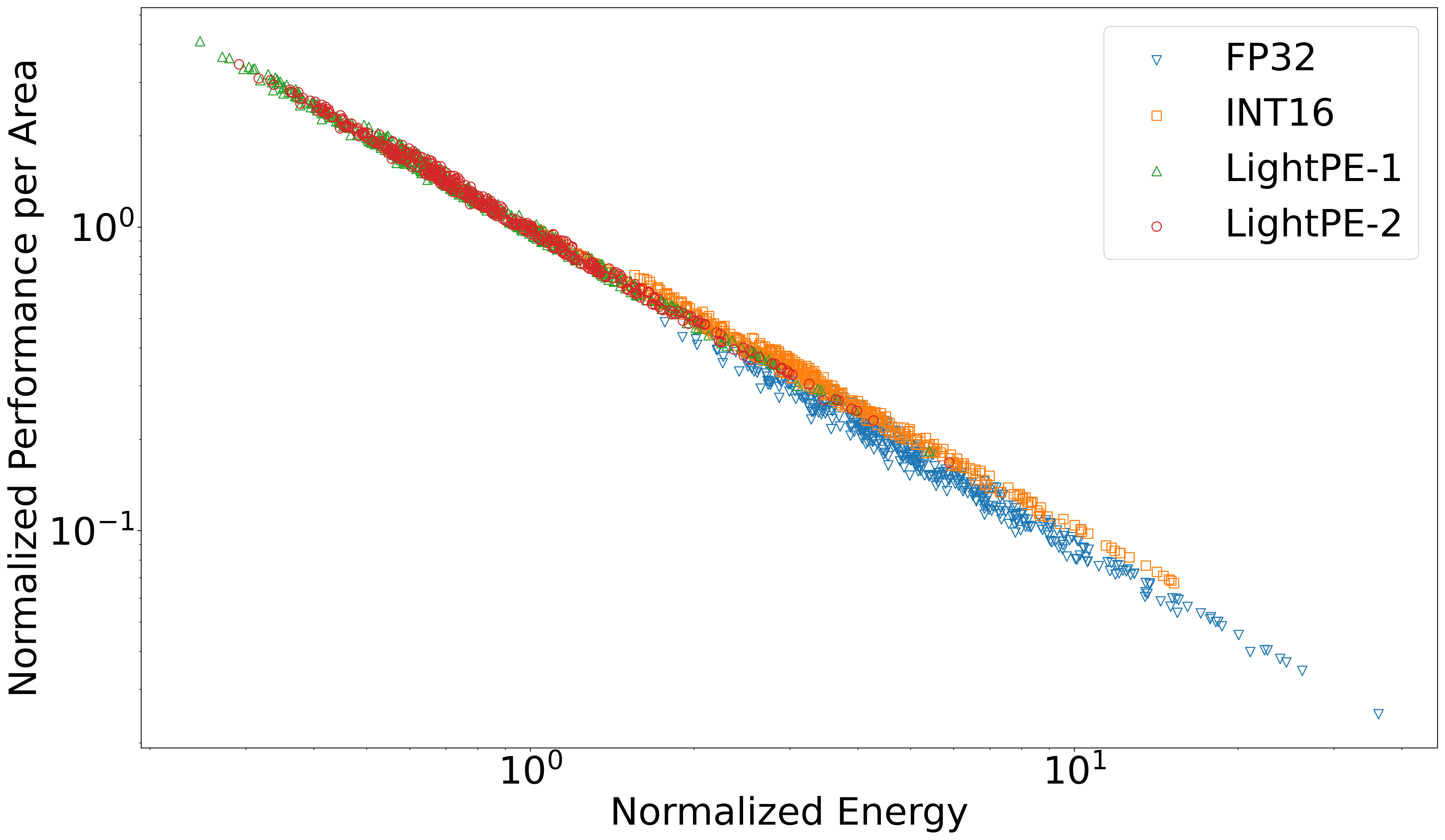}
  \includegraphics[width=0.32\textwidth]{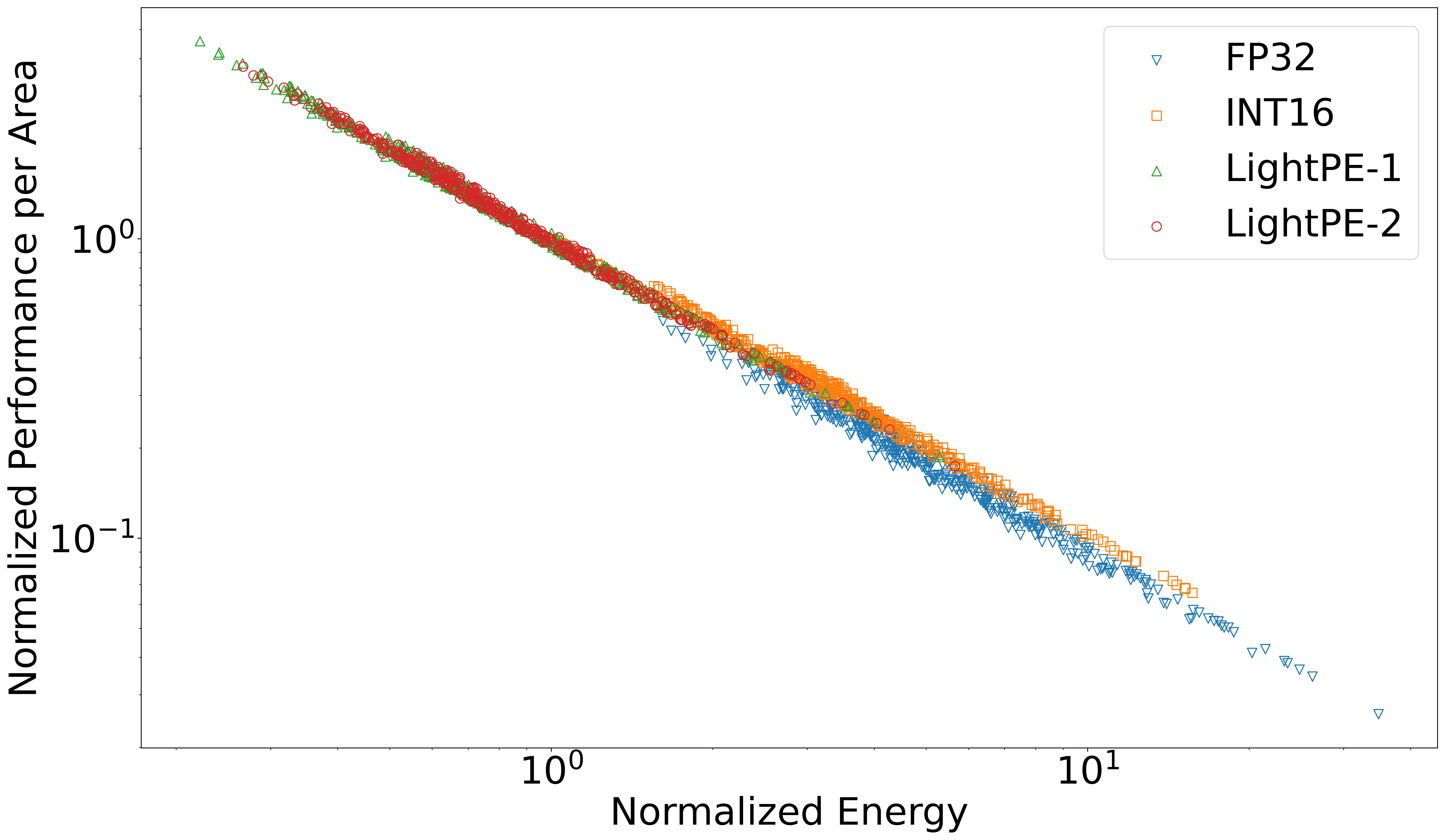}
    \\
 \includegraphics[width=0.32\textwidth]{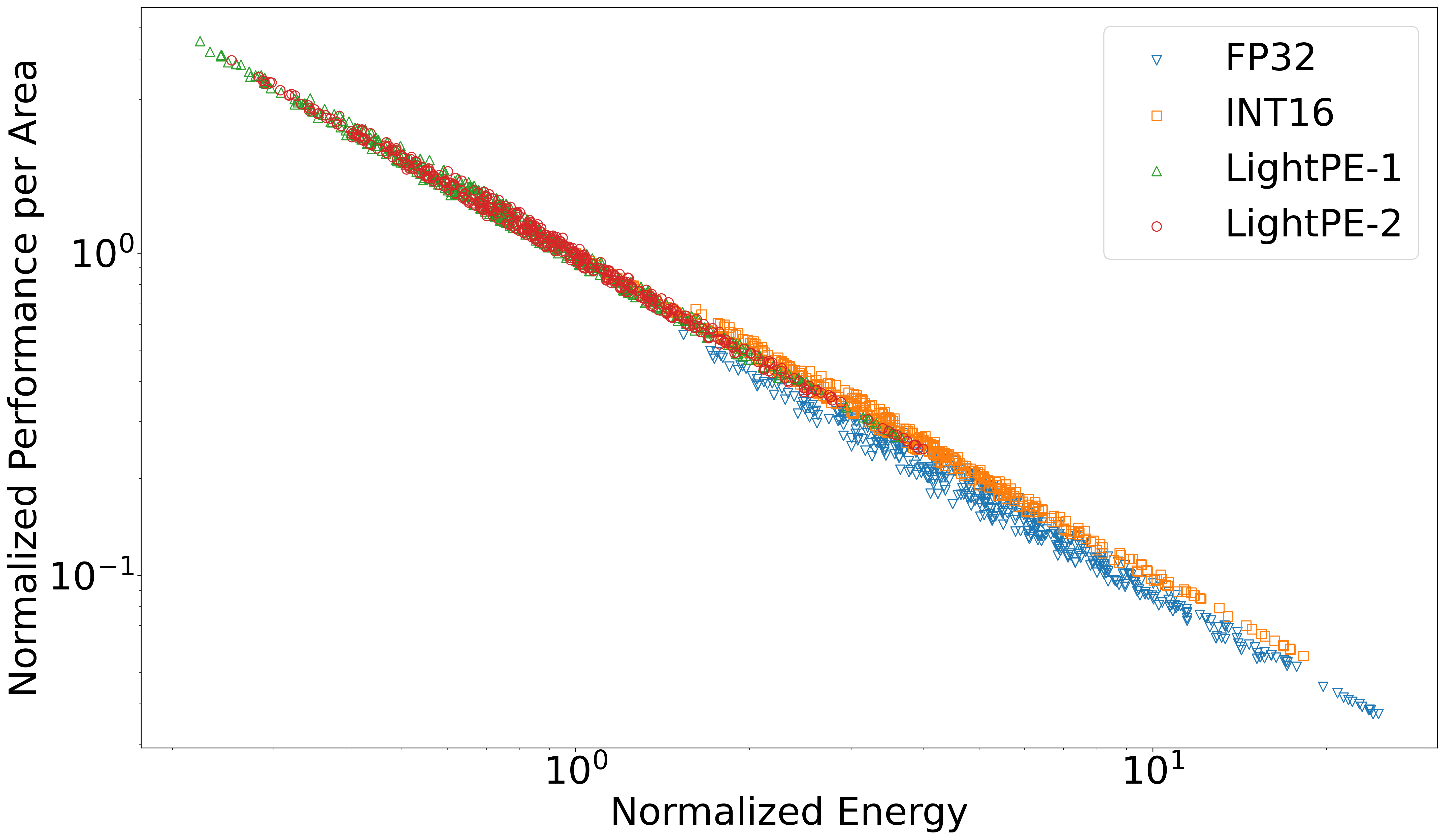}
  \includegraphics[width=0.32\textwidth]{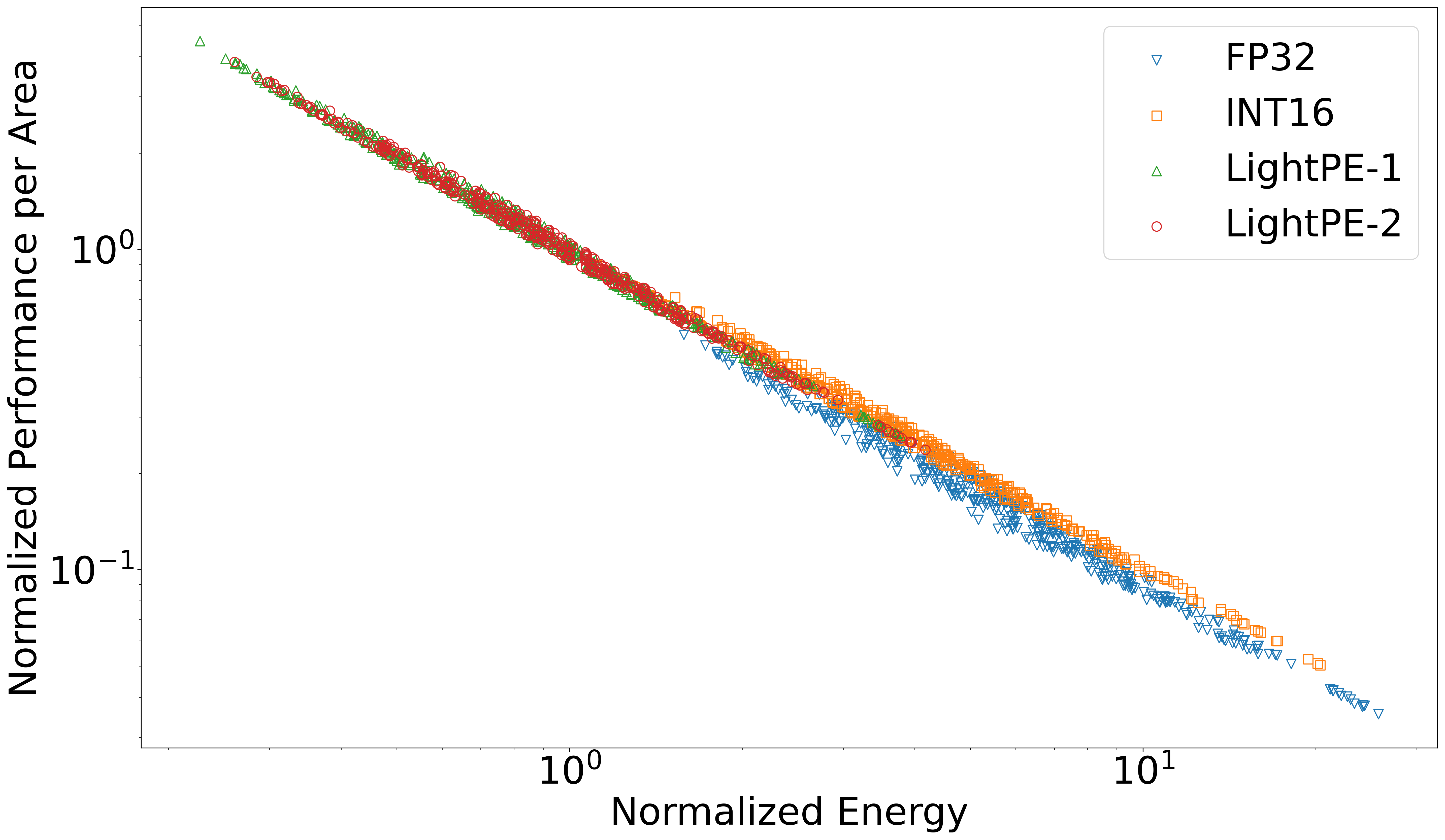}
  \includegraphics[width=0.32\textwidth]{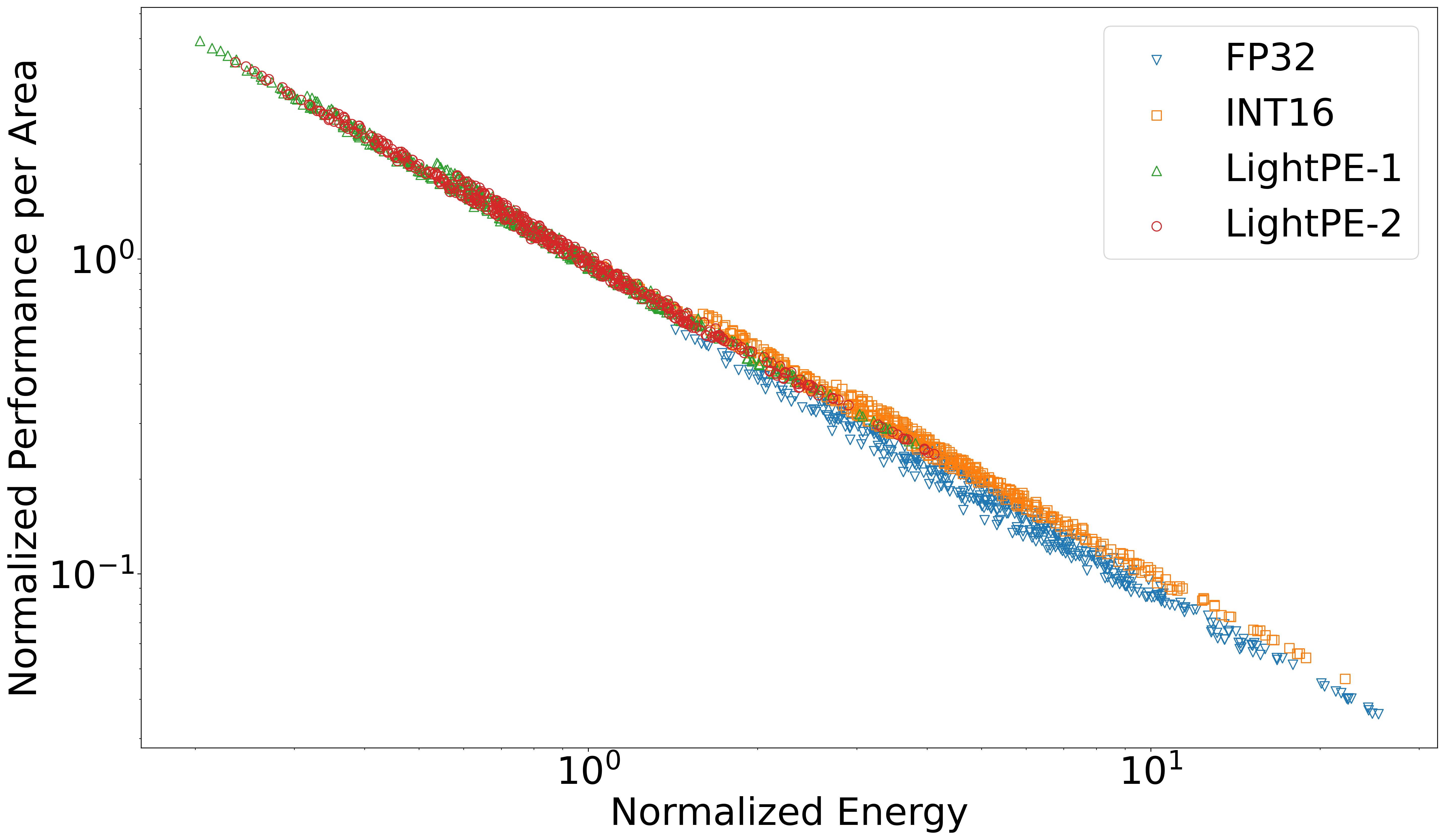}

 \caption{{Normalized performance per area vs. normalized energy results with respect to the INT16 hardware configuration with the highest performance per area for VGG-16 (left), ResNet-20 (middle), and ResNet-56 (right) for CIFAR-10 (top row), CIFAR-100 (middle row), and ImageNet (bottom row) design spaces. As it can be seen, LightPEs consistently outperform conventional INT16 and FP32 based designs in various models and datasets, thereby showing the benefits of using lower precision generalize across a variety of models.}}
  \vspace{-2mm}
 \label{fig:dse}
\end{figure*}

In this section, we present power, performance, and area modeling results for each processing element type and perform a design space exploration on various DNN models such as VGG-16 \cite{vgg16}, ResNet-20, ResNet-34, ResNet-50, and ResNet-56 \cite{resnet50} on CIFAR-10, CIFAR-100, and ImageNet datasets to iterate through our framework to demonstrate the flexibility of \textit{QADAM} for future studies. 

As detailed in Section III, \textit{QADAM} framework provides power, performance, and area models that significantly speed up the design space exploration. Figure~\ref{fig:model} shows the actual and estimated power, performance, and area results for each processing element type such as FP32, INT16, LightPE-1, and LightPE-2. Each data point in Figure~\ref{fig:model} corresponds to a different hardware accelerator configuration in the comprehensive design space. As shown by the results, \textit{QADAM}'s PPA models achieve high correlation to the actual PPA values. 
Figure~\ref{fig:model} also shows that the FP32 implementation has the highest area and power cost whereas LightPEs have the lowest area and power results when one processing element is considered. This shows the hardware-efficiency of LightPEs when compared to conventional PE implementations.


\subsection{Design Space Exploration Results}


To show the hardware-efficiency of LightPEs to conventional PE types, we perform design space exploration on VGG-16 \cite{vgg16}, ResNet-20, and ResNet-56 models on CIFAR-10/CIFAR-100 and VGG-16, ResNet-34, and ResNet-50 \cite{resnet50} models on ImageNet datasets as shown in Figure~\ref{fig:dse}. We show the normalized performance per area and normalized energy results for each PE type with respect to the baseline INT16 based implementation with the highest performance per area for the given design space. 

Figure~\ref{fig:dse} shows that LightPE implementations consistently outperform INT16 and FP32 implementations in both aspects, which proves their efficacy in terms of hardware-efficiency. Specifically, LightPE-1 and LightPE-2 achieve $4.8 \times$ and $4.1 \times$ more performance per area and $4.7 \times$ and $4 \times$ less energy on average across all workloads and datasets when compared to the best INT16 hardware configuration, respectively. On the other hand, INT16 baseline implementation achieves  $1.8 \times$ more performance per area and $1.5 \times$ less energy on average when compared to the best FP32 configuration. 

These conclusions hold for all the models and the datasets considered in this work such as VGG-16, ResNet-20, ResNet-34, ResNet-50, and ResNet-56 thereby showing that the benefits of using lower precision generalize across a variety of models. 
We conclude that different bit precisions and PE types can lead to significantly different performance per area and energy results which are two critical metrics for machine learning and systems community strives to improve upon.

\subsection{Pareto-Optimality for Accuracy and Performance per Area}

\begin{figure*}[h]
  \centering
 \includegraphics[width=0.49\textwidth]{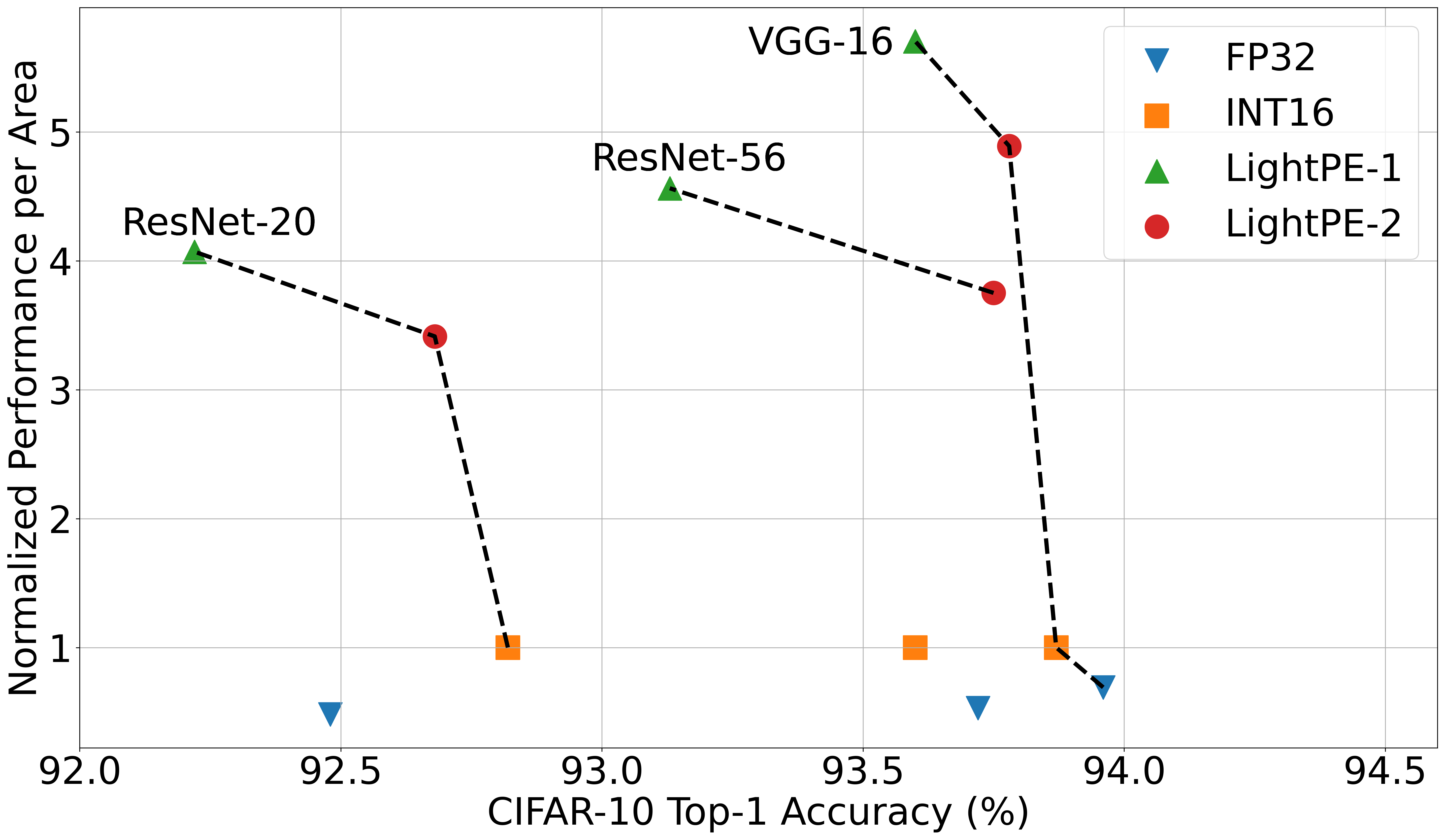}
  \includegraphics[width=0.49\textwidth]{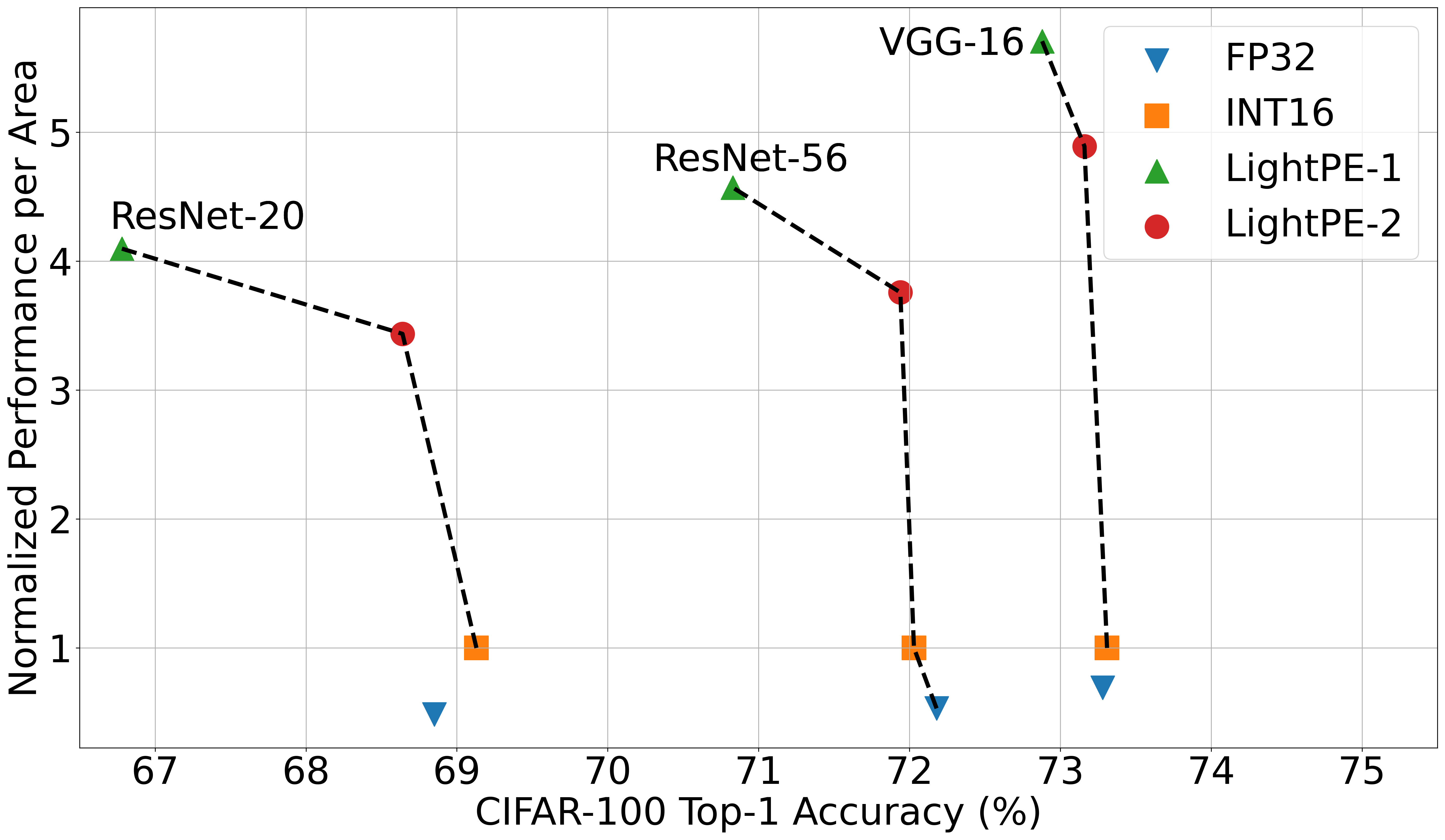}
 \caption{Normalized performance per area and top-1 accuracy results for various processing element types such as FP32, INT16, LightPE-1, and LightPE-2 for CIFAR-10 (left chart) and CIFAR-100 (right chart). Each data point corresponds to the hardware configuration with the highest performance per area for the corresponding processing element type. Pareto-front is shown with a dashed line for each DNN model. LightPEs are consistently on Pareto-front for various DNN models.}
  \vspace{-3mm}
 \label{fig:pareto_perf_area}
\end{figure*}

\begin{figure*}[h]
  \centering
 \includegraphics[width=0.49\textwidth]{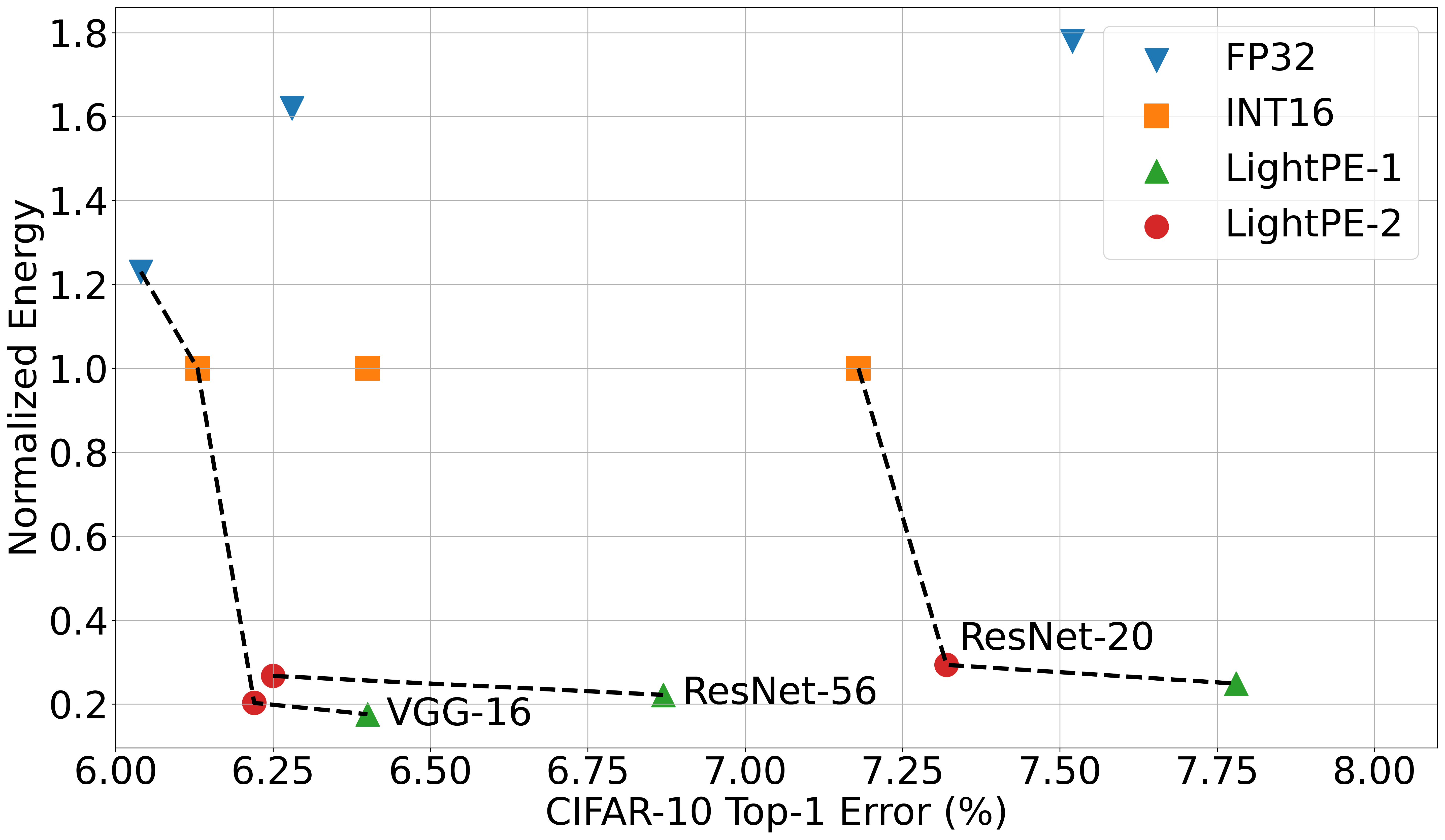}
  \includegraphics[width=0.49\textwidth]{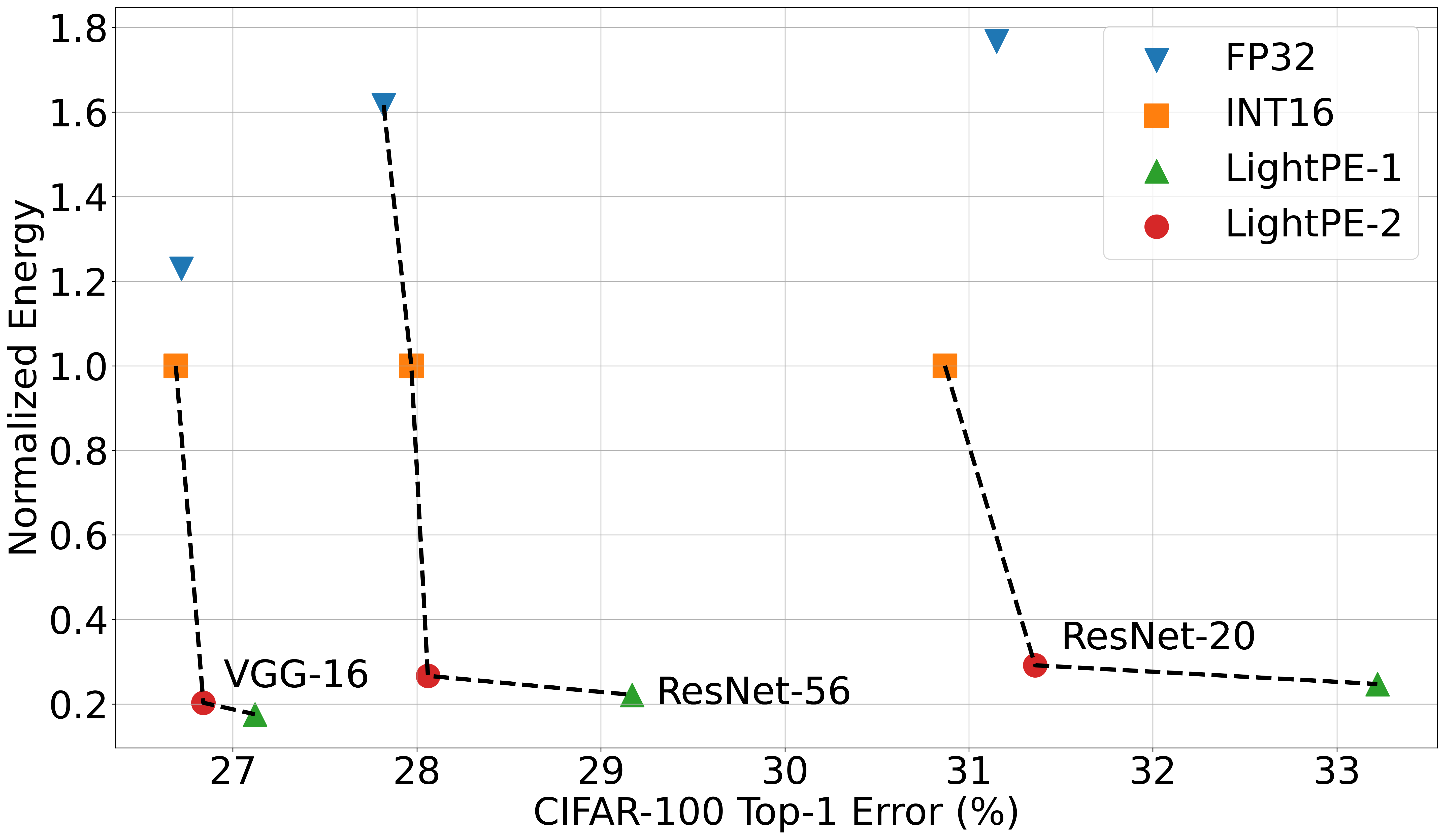}
 \caption{Normalized energy and top-1 error results for various processing element types such as FP32, INT16, LightPE-1, and LightPE-2 for CIFAR-10 (left chart) and CIFAR-100 (right chart). Each data point corresponds to the hardware configuration with the lowest energy for the corresponding processing element type. Pareto-front is shown with a dashed line for each DNN model. LightPEs are consistently on Pareto-front for various DNN models.}
 \vspace{-3mm}
 \label{fig:pareto_energy}
\end{figure*}

To show the accuracy and performance per area trade-off for different processing element types, we perform a Pareto-front analysis by training VGG-16, ResNet-20, and ResNet-56 models for CIFAR-10 and CIFAR-100 datasets. For both datasets, we perform five trials for each DNN model and processing element type and plot the mean top-1 accuracy results. The training recipe for both CIFAR-10/CIFAR-100 datasets follows prior art \cite{yang2018soft,Chin2020CVPR} which uses stochastic gradient descent with nesterov momentum, weight decay 0.0005, batch size 128, 0.1 initial learning rate with decrease by $5 \times$ at epochs 60, 120, and 160, and train for 200 epochs in total. We note that this training recipe is tuned for full-precision models. Therefore, the accuracy results for LightPE variants might be higher with proper hyperparameter tuning. 

Figure~\ref{fig:pareto_perf_area} shows the normalized performance per area and accuracy results for FP32, INT16, LightPE-1, and LightPE-2. Performance per area results are normalized with respect to the best INT16 configuration for each DNN model. We plot the hardware configurations with the highest performance per area results for each processing element type. Next, we perform a Pareto-front analysis among different processing element types and show the Pareto-frontier with a dashed line for each DNN model. We show that LightPEs are consistently on Pareto-front for various DNN models and datasets, whereas FP32 and INT16 based designs are occasionally dominated by LightPE variants. We show that LightPE-1 and LightPE-2 achieve on par accuracy results with FP32 and INT16 while achieving up to $5.7 \times$ and $4.9 \times$ more performance per area when compared to INT16 configuration, respectively.

\subsection{Pareto-Optimality for Accuracy and Energy}

We also perform a Pareto-front analysis for accuracy and energy results. We follow the same training methodology explained in Section IV-B. Figure~\ref{fig:pareto_energy} shows the normalized energy and accuracy results for FP32, INT16, LightPE-1, and LightPE-2 based designs. Energy results are normalized with respect to the best INT16 configuration for each DNN model. We show that LightPEs are systematically on Pareto-front for various DNN models and datasets. Specifically, LightPE-1 and LightPE-2 achieve $4.7 \times$ and $4 \times$ less energy on average across different workloads and datasets when compared to INT16 configuration, respectively. We also show that as model complexity increases, the accuracy gap between LightPEs and FP32 and INT16 based designs decreases. Thus, we conclude that our proposed LightPEs have promising results for training larger models with negligible accuracy loss while achieving significant performance per area and energy improvements. 

\vspace{-1mm}

\section{Conclusion}
\vspace{-2mm}
In this work, we present \textit{QADAM}, a quantization-aware highly parameterized power, performance, and area modeling framework for DNN accelerators. Our framework can foster the future research on design space exploration of DNN accelerators for various design choices such as bit precision, processing element type, scratchpad size of processing elements, global buffer size, device bandwidth, number of total processing elements in the design, and DNN workloads. 
Our results show that different bit precisions and processing element types lead to significant differences in terms of performance per area and energy. Specifically, LightPE-1 and LightPE-2 achieve $4.8 \times$ and $4.1 \times$ more performance per area and $4.7\times$ and $4 \times$ energy improvement on average when compared to the best INT16 hardware configuration, respectively. We also show that our proposed LightPEs consistently achieve Pareto-optimal results in terms of accuracy and performance per area and energy. Therefore, design space exploration of quantization-aware DNN accelerators merits a meticulous analysis that take these factors into account.

\section*{Acknowledgements}
This research was supported in part by National Science Foundation grants CCF No. 1815899 and CSR No. 1815780.


\bibliographystyle{IEEEtranS}
\bibliography{refs}

\end{document}